\documentclass[journal,draftcls,onecolumn,12pt,twoside]{IEEEtran}
\pdfoutput=1
\ifCLASSINFOpdf
\usepackage[pdftex]{graphicx}
\else
\fi
\usepackage{mathtools}
\usepackage{cite}
\usepackage{amsmath,amssymb,amsfonts}
\usepackage{algorithmic}
\usepackage{graphicx}
\usepackage{textcomp}
\usepackage{xcolor}
\usepackage{caption}
\usepackage{url}
\usepackage{epstopdf}
\usepackage{float}
\usepackage{booktabs}

\usepackage{epstopdf}
\usepackage{amsmath,amsthm, amssymb}
\usepackage{threeparttable}
\usepackage{bm}
\usepackage{multirow}
\usepackage{booktabs}
\usepackage{color}
\usepackage{tabularx}
\usepackage{setspace}
\usepackage{verbatim}
\usepackage{stfloats}

\usepackage{subfigure}
\usepackage{caption}
\usepackage{amsmath}

\usepackage{algorithm}
\usepackage{algorithmic}

\usepackage{lineno}
\begin{document}

\title{Time-Dependent Performance Modeling for Platooning Communications at Intersection}

\author{Qiong Wu,~\IEEEmembership{Member,~IEEE}, Yu Zhao and Qiang Fan
\thanks{This work was supported in part by the National Natural Science Foundation of China under Grant No. 61701197, in part by the open research fund of State Key Laboratory of Integrated Services Networks under Grant No. ISN23-11,  in part by the 111 Project under Grant No. B12018. \emph{(Corresponding author: Qiong Wu)}

Qiong Wu and Yu Zhao are with School of Internet of Things Engineering, Jiangnan University, Wuxi 214122, China, and also with the State Key Laboratory of Integrated Services Networks (Xidian University),  Xi'an 710071, China (Email: qiongwu@jiangnan.edu.cn, yuzhao@stu.jiangnan.edu.cn).

Qiang Fan is with Qualcomm, San Jose CA 95110 USA (e-mail: qiangfan29@gmail.com).
}
}

\markboth{}
{}

\maketitle

\begin{abstract}
With the development of internet of vehicles, platooning strategy has been widely studied as the potential approach to ensure the safety of autonomous driving. Vehicles in the form of platoon adopt 802.11p to exchange messages through vehicle to vehicle (V2V) communications. When multiple platoons arrive at an intersection, the leader vehicle of each platoon adjusts its movement characteristics to ensure that it can cross the intersection and thus the following vehicles have to adjust their movement characteristics accordingly. In this case, the time-varying connectivity among vehicles leads to the significant non-stationary performance change in platooning communications,  which may incur safety issues. In this paper, we construct the time-dependent model to evaluate the platooning communication performance at the intersection based on the initial movement characteristics. We first consider the movement behaviors of vehicles at the intersection including turning, accelerating, decelerating and stopping as well as the periodic change of traffic lights to construct movement model, and then establish a hearing network to reflect the time-varying connectivity among vehicles. Afterwards, we adopt the pointwise stationary fluid flow approximation (PSFFA) to model the non-stationary behavior of transmission queue. Then, we consider four access categories (ACs) and continuous backoff freezing of 802.11p to construct the models to describe the time-dependent access process of 802.11p. Finally, based on the time-dependent model, the packet transmission delay and packet delivery ratio are derived. The accuracy of our proposed model is verified by comparing the simulation results with analytical results.
\end{abstract}

\IEEEpeerreviewmaketitle
\section{Introduction}
\label{sec1}
\IEEEPARstart{A}{s} the number of vehicles keeps increasing rapidly, road safety remains to be a critical issue. The world health organization (WHO) reports that around 1.25 million people die from traffic accidents each year in the world \cite{Wang2018}. With the advent of artificial intelligence and cloud computing, autonomous driving has attracted the attention of many researchers due to its potential in improving road safety and travel experience \cite{zhu2021} -\cite{Xu2018}. Platooning is an important strategy to manage autonomous vehicles on a common lane. A platoon is composed of a leader vehicle and several follower vehicles. The leader vehicle controls the movement characteristics of the platoon, e.g., velocity and acceleration. The follower vehicles are autonomous vehicles which follow the leader vehicle one by one. The critical task of the platoon is to maintain its stable formation, i.e., each vehicle in the platoon keeps driving with the similar velocity and intra-platoon spacing. A platoon with the stable formation can save fuel consumption and avoid traffic accidents with high probability \cite{Liuyang2019,nie2018}.

Each platoon adopts the cooperative adaptive cruise control (CACC) system to maintain the stable formation, which relies on onboard sensors. Specifically, each vehicle in the platoon is equipped with sensors, e.g., radar or stereo cameras, to distinguish the preceding vehicle in the platoon from the nearby vehicles and detect the distance toward the preceding vehicle \cite{Widmann2000,xia2018}. Once the detected distance is not the safety intra-platoon spacing, the vehicle will automatically adjust its velocity and acceleration to ensure the distance to be the safe value \cite{ChenPark2020}. However, sensors may be adversely affected by their own hardware problems and the external environment, e.g., rainy and foggy weather \cite{Ahmed-Zaid}, which may prevent the vehicle from exactly distinguishing the preceding vehicle from the nearby vehicles. In this case, these sensors may fail in the target detection and yield an inaccurate detected distance \cite{Austin1987}, so it is far from enough to solely rely on the onboard sensor to maintain the stable platoon formation. Therefore, the CACC system further utilizes vehicle to vehicle (V2V) communications to maintain the stable platoon formation. IEEE 802.11p and cellular-vehicle to everything (C-V2X) Mode 4 are two important technologies to realize V2V communications without any infrastructure support \cite{Segata}. However, C-V2X Mode 4 incurs higher average delay than 802.11p \cite{Discrete-Time2021}. Meanwhile, C-V2X Mode 4 lacks the mechanism to support the four kinds of data streams with different priorities defined by the european telecommunications standard institute (ETSI) \cite{Wijesiri}, while 802.11p uses the enhanced distributed coordination function (EDCA) mechanism at medium access control (MAC) layer to support the messages of the four access categories (ACs), where the ACs of messages are classified according to safety levels of vehicles \cite{Zheng2016}. Thus, C-V2X Mode 4 is limited to support V2V communications for the autonomous vehicles in platoons \cite{Segata2021}. Therefore, in this paper we consider each vehicle adopts 802.11p to communicate with each other. Specifically, each vehicle exchanges messages with the vehicles within its communication range\cite{1809.02867}. The messages have to be received successfully within a limited time interval to ensure that all vehicles in the platoon can react in time to maintain the stable platoon formation. Therefore, the packet transmission delay (PTD) and packet delivery ratio (PDR) are very important metrics for platooning communications.


The characteristics of vehicles in platoons change dynamically when the platoons drive across the intersection. Specifically, when a platoon enters an intersection area, the leader vehicle will accelerate to pass through the intersection or decelerate to stop before the stop line according to the status of the traffic light, i.e., red or green, which incurs dramatic change of  intra-platoon spacing toward its follower vehicles. In this case, the follower vehicles of the platoon will adjust their acceleration to maintain the intra-platoon spacing between adjacent vehicles, thus the formation of the platoon will be changed dynamically. Especially, when the platoon is turning at the intersection, the moving direction of the leader vehicle is time-varying. In this case, each follower vehicle will sense the varying distance toward the preceding vehicle and adjust its moving direction and acceleration accordingly to maintain the safe intra-platoon spacing. In addition, when the distance between two adjacent platoons, i.e., inter-platoon spacing, is smaller than the safe intra-platoon spacing, the leader vehicle of the following platoon will also adjust its acceleration to ensure the safe spacing, thus the formation of the following platoon will also change dynamically. As a result, when multiple platoons on different lanes with different directions enter the intersection, the distances between vehicles change obviously, which imposes the network connectivity to change dynamically. In this case, the PTD and PDR will be time-dependent, thus the vehicles in the platoons may be unable to receive messages successfully within the limited time duration and further react in time to maintain the stable formation. It is critical to establish an analytical time-dependent model to evaluate the complex system scenario of vehicular platooning. However, the dynamic network connectivity leads to the non-stationary change of the PTD and PDR of platooning communications, which poses a challenge to establish the time-dependent model. To the best knowledge of us, no work has constructed the time-dependent model to evaluate the performance of platooning communications in the complex intersection scenario, which motivates us to investigate it.

In this paper, we construct a time-dependent model to evaluate the performance of vehicular platooning communications at an intersection. The main contribution of this paper is summarized as follows.

\begin{itemize}
\item[1)] Considering the movement behaviors of platoons at the intersection (including turning, accelerating, decelerating and stopping) as well as the periodic change of traffic lights, we construct a movement model of platoons at the intersection to calculate the time-dependent position of each vehicle. We further construct a hearing network to calculate the time-varying connectivity among vehicles according to the vehicles' positions.
\item[2)] The pointwise stationary fluid flow approximation (PSFFA) is adopted to model the non-stationary dynamic behavior of the transmission queue. Moreover, we consider the four ACs and continuous backoff freezing of 802.11p to construct the models to describe the time-dependent access process of 802.11p. Then, the PTD and PDR are derived based on our proposed time-dependent model.
\item[3)] The accuracy of our proposed model is verified by comparing the analytical results with the simulation results. Moreover, the time-dependent performance of platooning communications at the intersection has been evaluated based on our model.

\end{itemize}

The rest of the paper is organized as follows. Section \ref{sec2} reviews related works on the performance modeling of V2V communications. Section \ref{sec3} describes the system scenario and overviews the 802.11p EDCA mechanism. Section \ref{sec4} constructs the movement model to derive the time-dependent position of each vehicle and then establishes the hearing network to reflect the time-varying connectivity among vehicles. Section \ref{sec5} constructs time-dependent model for platooning communications at the intersection and then derives the PTD and PDR. We present simulation and analytical results in Section \ref{sec6}, and then conclude them in Section \ref{sec7}.

\section{Related Work}
\label{sec2}
In this section, we have reviewed the existing works on the performance modeling of V2V communications including 802.11p and C-V2X Mode 4.

Many works have focused on the performance modeling of 802.11p and C-V2X Mode 4 in vehicular ad hoc networks (VANETs).
In \cite{Kim2017}, Kim \textit{et al.} constructed multi-dimensional Markov chains to model the performance of 802.11p in intelligent transportation systems, and then evaluated the successful delivery probability and the delay distribution.
In \cite{Shahen2019}, Shahen \textit{et al.} presented Markov chain model to derive the throughput and average delay of the 802.11p, and further verified whether the 802.11p can satisfy the criteria of VANETs.
In \cite{Almohammedi2021}, Almohammedi \textit{et al.} derived the transmission probability and system throughput of 802.11p based on a non-saturated 2-D Markov chain model to evaluate the performance of vehicular networks.
In \cite{Yao2019}, Yao \textit{et al.} presented a performance analysis model to evaluate the probability distribution, deviation and mean of the MAC access delay for the single-hop broadcast under the 802.11p protocol.
In \cite{Togou2018}, Togou \textit{et al.} proposed a Markovian analytical model to describe the process of collecting time-dependent channel state information and analyzed the throughput of the 802.11p for unsafety applications in V2V communications.
In \cite{Zheng2016}, Zheng and Wu considered the factors (such as saturation condition, backoff counter freezing and standard parameters) that affect the transmission probability, collision probability and delay of 802.11p, and constructed a 1-D and a 2-D Markov chain to model the 802.11p.
In \cite{Gonzalez2019}, Gonzalez \textit{et al.} presented the models to analyze the average PDR of C-V2X Mode 4, and then quantified the four different types of packet errors that affect C-V2X Mode 4.
In \cite{Schiegg2019}, Schiegg \textit{et al.} considered the physical (PHY) and MAC layer of C-V2X Mode 4, while developed a comprehensive analytical model to evaluate the communication performance.
Although the above works have studied the modeling of the 802.11p and C-V2X Mode 4 in VANETs, they have not considered the platooning scenario.

Some existing works have studied the performance modeling of 802.11p and C-V2X Mode 4 for platooning communications.
In \cite{Peng2017}, Peng \textit{et al.} presented a 802.11p-based communication model in multi-platooning scenario to improve the effectiveness of information sharing for platoon controlling.
In \cite{Jornod2020}, Jornod \textit{et al.} constructed a prediction model to guarantee the packet inter-reception time in the 802.11p-based VANETs where the vehicles cooperate as a platoon.
In \cite{Thunberg2019}, Thunberg \textit{et al.} proposed an analytical framework considering the characteristics of V2V communication, i.e., packet loss probabilities and packet transmission delays, to guarantee the string stability of platoon in the 802.11p-based VANETs.
In \cite{Yu2018}, Yu \textit{et al.} adopted the Markov process and M/G/1/K queuing theory to propose the platoon structure model, vehicle control model and communication model in a single platoon scenario and demonstrated that the 802.11p-based intra-vehicle communication can guarantee the stability of platoon.
In \cite{Lekidis}, Lekidis \textit{et al.} presented a novel C-V2X Mode 4 network framework for platooning applications and constructed models to demonstrate the improvement of computing latency.
In \cite{Fu2021}, Fu \textit{et al.} considered the C-V2X Mode 4 and proposed a prediction-assisted platooning mechanism to optimize the communication performance of platoons. They constructed model for each vehicle based on the information received from the platoons to reduce information latency, and then designed a detection algorithm to protect the platoons from malicious vehicles.
However, these works only studied the time-independent performance model of vehicular platooning communications, which is not suitable to evaluate the dynamic vehicular environment when multiple platoons are crossing the intersection area.


A few papers have presented time-dependent performance model of vehicular communications in VANETs.
In \cite{Xu2016}, Xu \textit{et al.} established a time-varying communication connectivity network and evaluated the 802.11p-based V2V communication network.
In \cite{Tong2016}, Tong \textit{et al.} adopted a continuous-time Markov chain to investigate the transmission behavior of 802.11p based V2V safety communications.
In \cite{Awad2019}, Awad \textit{et al.} considered the impact of time-varying channel on V2V communications, and proposed training-based, blind and semi-blind algorithms to evaluate the 802.11p-based V2V communications.
In \cite{Ge2020}, Wu \textit{et al.} considered the disturbance in platoons and established a time-varying communication connectivity network to evaluate the PTD and PDR of the 802.11p in platoons.
However, these works constructed the time-dependent model of vehicular communications in VANETs without considering platooning communications at the intersection, which inspires us to conduct this work.

\begin{table}\tiny
\caption{Notations used in this section}
\scriptsize
\begin{center}
\begin{tabular}{p{50pt}p{180pt}}
\hline
$A_m$ & the number of additional time slots that must be waited to detect the channel idle for $AC_m$ compared with $AC_0$  \\
$AIFSN_m$  & the Arbitration Inter Frame Space Number of $AC_m$  \\
$a$  &  the maximum acceleration in the IDM model  \\
$a_{k,i}(t)$ & the acceleration of $V_{k,i}$ at time $t$  \\
$b$   &   the suitable deceleration in the IDM model  \\
${B_j^{k,i,m}}(z)$    & the PGF of the time that the backoff counter of $AC_m$ in vehicle $V_{k,i}$ decreases to 0 when the number of retransmissions is $j$   \\
$c_{k,i,m}^2(t)$   &   the squared coefficient variation of service time  \\
$D$  &   the distance from the border of intersection area to stop line  \\
$D_r$   &   the coverage of RSU  \\
$D_s$   &   the distance from the stop line to the center line  \\
$E[P]$ & the packet size  \\
${F_m}(z)$  &   the PGF of the backoff frozen time  \\
$f_s^{k,i,m}(t)$ &  the number of successfully received packets for $AC_m$ of $V_{k,i}$ at time $t$ \\
$f_a^{k,i,m}(t)$ & the number of packets arriving at $AC_m$ of $V_{k,i}$ at time $t$ \\
${H_{k,i,m}}(z)$   &   the PGF of the time that the backoff counter of $AC_m$ in vehicle $V_{k,i}$ decreases one \\
$L_0$    &   the length of each vehicle  \\
$MAC_{H}$ & the header length of MAC layer \\
$M_m$  &  the maximum number of retransmissions that the contention can be doubled  \\
$M_m^l$ &  the retransmission limit  \\
$N_{c}^{k,i}(t)$   &  the total number of vehicles in the communication range of $V_{k,i}$ at time $t$  \\
$N_{k,i,m}(t)$   &  the average number of packets in $AC_m$ queue at time $t$  \\
$\dot{N}_{k,i,m}(t)$  &  the change rate of $N_{k,i,m}(t)$  \\
$n_k$ & the number of vehicles of platoon $P_k$\\
$P$ & the number of platoons\\
$P_k$   &  the $k$th platoon  \\
$PT{D_{k,i,m}}(t)$ & the time-dependent packet transmission delay  \\
$PDR_{k,i,m}(t)$  & the time-dependent packet delivery ratio \\
$P_{{\rm{st}}}^{k,i,m}(z)$   &   the PGF of service time of $V_{k,i}$'s $AC_m$  \\
$PHY_{H}$ &  the header length of physical layer  \\
$p_a^{k,i,m}(t)$   &   the probability that a packet arrives at the $AC_m$ queue of vehicle $V_{k,i}$  \\
$p_b^{k,i,m}(t)$   &   the backoff freezing probability of $AC_m$  in vehicle $V_{k,i}$ \\
$p_v^{k,i,m}(t)$   & the internal collision probability of $AC_m$ that multiple $ACs$ including $AC_m$ transmit packets at the same time.  \\
${p_{{ki},{ql}}^s}(t)$ & the probability that the vehicle $V_{q,l}$ successfully receives a packet transmitted from the target vehicle $V_{k,i}$ \\
${p_{{ki},{ql}}^{exposed}}(t)$  &  the collision probability caused by exposed vehicles \\
${p_{{ki},{ql}}^{hidden}}(t)$  &  the collision probability caused by hidden vehicles  \\
$R_l$   &  the left-turn radius  \\
$R_r$   &  the right-turn radius  \\
$R_b$ & the basic rate \\
$R_c$ & the communication range of each vehicle \\
$R_d$ & the data rate  \\
$s_0$  &  the minimum intra-platoon spacing \\
$s_e$  &   the intra-platoon spacing at the equilibrium point $e$ in the IDM model  \\
$\Delta S_{k,i}(t)$    &     the intra-platoon spacing between $V_{k,i}$ and $V_{k,i-1}$ at time $t$   \\
$\Delta L_{k,i}(t)$ & the driving distance of $V_{k,i}$ from time $t$ to time $t+\Delta t$   \\
$S_{k,i}^*(t)$        &     the desired gap of $V_{k,i}$ to $V_{k,i-1}$  at time $t$ \\
$T_0$   &   the desired time headway  \\
$T_G$   & the time duration of the green traffic light  \\
$T_R$   & the time duration of the red traffic light  \\
$T_G^{rem}$   &  the remaining time of green light when the leader vehicle enters the intersection area \\
${T_{tr}}(z)$    &   the PGF of the transmission time  \\
$T_{k,i,m}(t)$   &  the MAC service time of $AC_m$ in vehicle $V_{k,i}$  \\
$T_{k,i,m}^s(t)$ & the average residence time of a packet in $AC_m$'s queue  \\
$v_0$ &    the maximum velocity allowed on the road \\
$v_e$     &  the velocity at the equilibrium point $e$  \\
$v_{k,i}(t)$    &  the velocity of $V_{k,i}$ at time $t$    \\
$\Delta v_{k,i}(t)$    &   the velocity difference between $V_{k,i}$ and $V_{k,i-1}$   \\
$V_{k,i}$   &   the target vehicle $i$ in the platoon $k$   \\
$W_{m,j}$   &  the contention window size of $AC_m$ when the number of retransmission is $j$  \\
${w_{k,i,m}}(t)$  & the internal transmission probability of $AC_m$   \\
$(x_{k,i}(t),y_{k,i}(t))$    &   the position of vehicle $V_{k,i}$ at time $t$   \\
$\theta _{k,i}(t)$    &  the angle of $V_{k,i}$'s heading deviating from the X-axis  \\
$\sigma _{k,i,m}^2(t)$ & the variance of the service time in $AC_m$ of $V_{k,i}$ \\
$\lambda _{m}$  &  the packet arrival rate at the $AC_m$ queue\\
$\mu _{k,i,m}(t)$   &    the service rate of $AC_m$  \\
${\rho _{k,i,m}}(t)$  &   the probability of non-empty transmission queue of $AC_m$ in vehicle $V_{k,i}$ \\
$\tau _{k,i}(t)$   &  the transmission probability of vehicle $V_{k,i}$ \\
${\tau _{k,i,m}}(t)$ & the external transmission probability of $AC_m$  \\
$\delta$ & the duration of one time slot  \\
$\gamma$ & the propagation delay  \\
\hline
\end{tabular}
\label{tab1}
\end{center}
\end{table}

\section{System Model}
\label{sec3}
In this section, we will introduce the system scenario at the intersection. Specifically, we describe the platoon behaviors at the intersection and then introduce the 802.11p EDCA mechanism. The notations in this paper are summarized in Table \ref{tab1}.

\subsection{Scenario of Multiple Platoons at Intersection}
\
\newline
\indent
 Consider that $P$ platoons are driving on a three-lane road at each direction of an intersection. The three lanes are left-hand lane, straight lane and right-hand lane, respectively. Each platoon obeys the traffic regulation, i.e., the platoon on the left-hand (right-hand) lane will turn left (right) at the intersection to its target lane, while the platoon on the straight lane will drive straightly and pass the intersection. The length of each stop lane is $2D_s$. The square area within the stop lanes is the center area of the intersection. Four traffic lights are deployed at the intersection to control the traffic flow of each direction. Each traffic light keeps green for duration $T_G$ and red for duration $T_R$. Consider the transceiver is placed at the front bumper of each vehicle. The leader vehicle of each platoon adjusts its movement according to the traffic light when it enters the intersection area, i.e., the distance of its front bumper toward the stop line is smaller than $D$, where $D$ is the distance from the border of intersection area to stop line.

\begin{figure}
\center
\includegraphics[width=8.6cm]{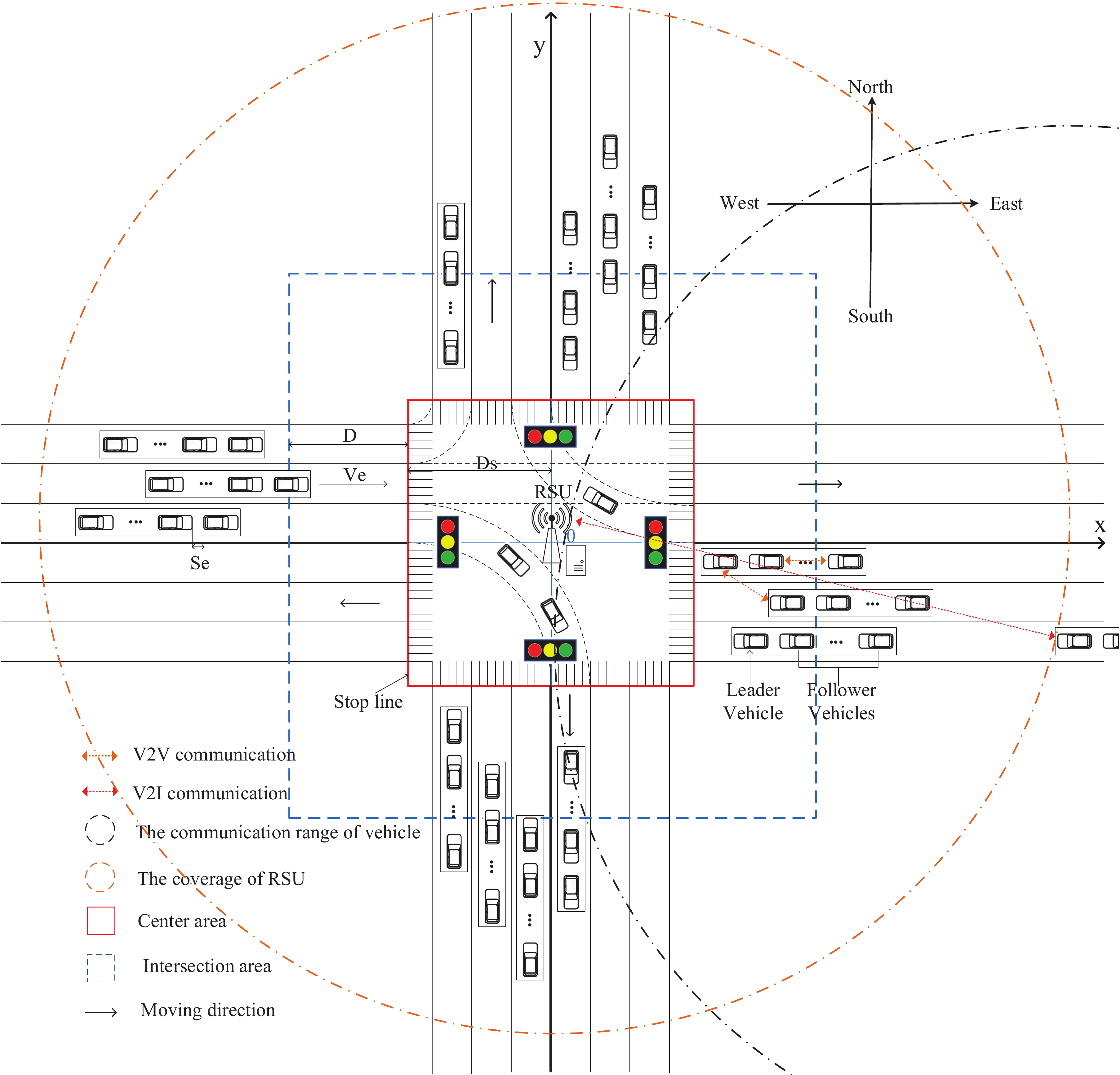}
\caption{Initial scenario}
\label{fig1}
\end{figure}

Within each time slot $\Delta t$, each vehicle in a platoon employs 802.11p to broadcast its movement characteristics, i.e., position, velocity and acceleration, to the vehicles within its communication range. A road-side unit (RSU) which is connected with a server is deployed at the center of the intersection to collect the initial movement characteristics of vehicles. Consider the coverage of the RSU is a circle with radius $D_r$, which is larger than $D+D_s$. The communication area of each vehicle is same with the coverage of the RSU. Each vehicle sends its initial movement characteristics to the RSU through vehicle to infrastructure (V2I) communications once it enters the coverage of the RSU, i.e., the bumper of the vehicle reaches the coverage of the RSU. The server connected with the RSU adopts our proposed model to predict the time-dependent position based on the received initial movement characteristics and further analyze the time-dependent communication performance for each vehicle. The scenario is shown in Fig. \ref{fig1}.

The movement of each leader vehicle is impacted by the inter-platoon spacing, i.e., the distance from the bumper of the leader vehicle to the bumper of the last vehicle in the preceding platoon. If the inter-platoon spacing is smaller than the safe intra-platoon spacing $s_e$, the leader vehicle would decelerate according to the intelligent driving model (IDM) which is a car-following model widely adopted by platoons until the inter-platoon spacing is resumed to $s_e$ \cite{Peng2017}. In contrast, if the inter-platoon spacing exceeds $s_e$, the leader vehicle's movement is dependent on different scenarios, i.e., 1) enter the coverage of the RSU; 2) enter the intersection area; 3) enter the center area of the intersection; 4) leave the center area of the intersection. Denote $P_k$ as the $k$th platoon, $V_{k,i}$ as the $i$th vehicle of platoon $P_k$ and $n_k$ as the number of vehicles of platoon $P_k$. Next, we will introduce the different behaviors of a leader vehicle $V_{k,1}$ of platoon $P_k$ in the four above cases.



\subsubsection{Enter the coverage of RSU}
\
\newline
\indent
When $V_{k,1}$ enters the coverage of the RSU, $P_k$ keeps a stable formation. $V_{k,1}$ does not change its direction and keeps moving with constant velocity $v_e$ until it enters the intersection area.

\subsubsection{Enter the intersection area}
\
\newline
\indent
When $V_{k,1}$ enters the intersection area, $V_{k,1}$ does not change its direction. It observes the traffic light and then changes the acceleration as follows until it enters the center area of the intersection.

\paragraph{red light}
\
\newline
\indent
$V_{k,1}$ makes the movement proposed in \cite{Kamal2015} to avoid emergency braking before stopping at the stop line. Specifically, $V_{k,1}$ reduces its acceleration in real time to slow down and stop until arriving at the stop line.


\paragraph{green light}
\
\newline
\indent
Similar with \cite{Smith2020}, we consider that a platoon is deemed to pass the stop line if the leader vehicle of the platoon passes the stop line. Thus, $V_{k,1}$ first observes the remaining time of green light and then estimates whether it can accelerate with the maximum acceleration $a$ specified in \cite{Peng2017} to ensure it passes the stop line before the due time of the green light \cite{Smith2020}. If $V_{k,1}$ can pass the stop line, $V_{k,1}$ makes a uniform acceleration linear motion with acceleration $a$ until it passes the stop line; Otherwise, it reduces its acceleration in real time, similar to its operation under the red light. Note that if its velocity exceeds the maximum velocity allowed on the road $v_0$, it would keep moving with the velocity $v_0$.

\subsubsection{Enter the center area of the intersection}
\
\newline
\indent
When $V_{k,1}$ enters the center area of the intersection, it would change the movements as follows until it leaves the center area of the intersection. If its velocity is larger than $v_e$, it decelerates its velocity with the suitable deceleration $b$ specified in \cite{Peng2017}. Otherwise, if its velocity is smaller than $v_e$, it accelerates its velocity with acceleration $a$. In the deceleration or acceleration process, if its velocity reaches $v_e$ it will keep the constant velocity.

Similar with \cite{Bichiou2019}, the moving directions of $V_{k,1}$ would be different when it moves on different lanes before entering the center area. If it's on the straight lane, it does not change its direction. If it's on the left-turn/right-turn lane, it takes a quarter circular motion until it reaches the target lane.

\subsubsection{Leave the center area of the intersection}
\
\newline
\indent
When $V_{k,1}$ leaves the center area of the intersection, it drives along a straight line and makes the same acceleration or deceleration process as it enters the center area of the intersection.

The follower vehicles in $P_k$ adjust their velocities according to the IDM model, thus the formation of the platoon is dynamic in different areas.

\begin{figure}
\center
\includegraphics[scale=0.45]{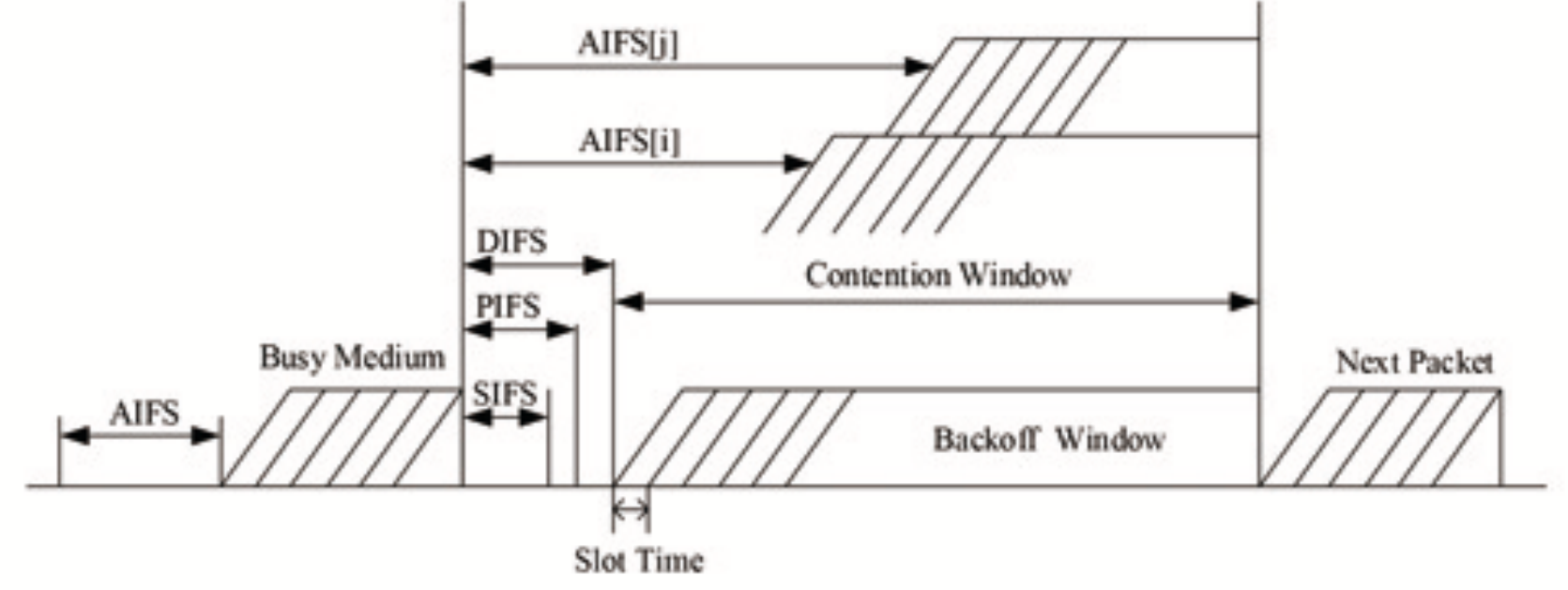}
\caption{Access process of the 802.11p EDCA mechanism}
\label{fig2}
\end{figure}

\subsection{The 802.11p EDCA Mechanism}
\
\newline
\indent
We will overview the access mechanism of the 802.11p protocol in this sub-section. The 802.11p protocol adopts the EDCA access mechanism at the MAC layer to satisfy the requirements of different services. It defines four ACs, i.e., $AC_m$ $(m=0,1,2,3)$, to support the packets with various priorities to access channel. Each AC has a separate transmission queue. To distinguish the priorities of different ACs, the 802.11p EDCA mechanism defines different contention parameters for each AC, including the minimum contention window $CW_{min}^m$, maximum contention window $CW_{max}^m$ and arbitration inter frame space $AIFS_m$. Here, $AIFS_m$ is a channel's idle duration before obtaining transmission opportunities, which is determined by arbitration inter frame space number $AIFSN_m$. The relationship between $AIFS_m$ and $AIFSN_m$ of $AC_m$ is expressed as

\begin{equation}
AIFS_m = AIFSN_m \times \delta  + SIFS,
\label{eq1}
\end{equation}where $SIFS$ is the short inter frame space and $\delta$ is one time slot.

In this paper, we adopt the broadcast mechanism of the 802.11p to access channel. The detailed access process is described as follows. The backoff process is initialized once $AC_m$ in a vehicle has to transmit a packet. Firstly, the backoff counter of $AC_m$ is set as a random value within $[0, W_{m,0}-1]$. Here, $W_{m,0}$ is the contention window size of $AC_m$ when the number of retransmissions is $0$ and $W_{m,0}=CW_{min}^m+1$. Then, if the channel is detected to be idle within one time slot $\delta$, the backoff counter of $AC_m$ would decrease by one. Otherwise, the backoff counter will be frozen until the channel keeps idle for the duration of $AIFS_m$. Note that continuous backoff freezing occurs when the channel is sensed to be busy for continuous slots. Therefore, the backoff counter would be frozen continuously until the channel keeps idle for $AIFS_m$. When the backoff counter of $AC_m$ decreases to 0, $AC_m$ attempts to transmit a packet. If no other $ACs$ with higher priorities in the vehicle are transmitting at the same time, the packet of $AC_m$ will be transmitted, and then the backoff counter will be reset as a random value within $[0, W_{m,0}-1]$. Otherwise, an internal collision occurs and the packet will be retransmitted. At this time, the number of retransmission pluses one and a new backoff process is initialized with the contention window $W_{m,1}$ to retransmit the packet. The relationship between $W_{m,j}$ $(j=0,1,\dots,M_m^l)$ and $CW_{min}^m$, $CW_{max}^m$ is expressed as

\begin{equation}
W_{m,j} = \left\{ \begin{array}{l}
 \begin{array}{*{20}{c}}
   {2^j}(CW_{min}^m + 1), {j \in [0,{M_m}]}  \\
\end{array} \\
 \begin{array}{*{20}{c}}
   CW_{max}^m+1, {j \in ({M_m},{M_m^l}]}  \\
\end{array} \\
 \end{array} \right.,
\label{eq2}
\end{equation}where $M_m$ is the maximum number of retransmissions that the contention window can be doubled, $M_m^l$ is the retransmission limit, and the $CW_{max}^m$ is calculated as

\begin{equation}
CW_{max}^m = 2^{M_m}(CW_{min}^m+1)-1.
\label{eq3}
\end{equation}

Let $M_m^l = M_m + L_m$. If the number of retransmissions exceeds $M_m^l$, the packet will be discarded. The detailed access process is shown in Fig. \ref{fig2}.

%

\section{Movement Modeling and Hearing network}
\label{sec4}
In this section, we first consider the movement behaviors of platoons at the intersection such as turning, accelerating, decelerating and stopping as well as the periodic change of traffic lights, to construct the movement model of vehicles and derive the time-dependent position of each vehicle. Then, we establish the hearing network according to the time-dependent position of each vehicle to reflect the time-varying connectivity among vehicles.


\subsection{Movement Modeling}

The position of each vehicle is represented by a rectangular coordinate system, where the center of the intersection is set as the origin, the direction of x-axis is east and the direction of y-axis is north. The RSU is located at the origin point. At the initial time $t_0$, $V_{k,1}$ arrives at the coverage of the RSU and sends the movement characteristics of each vehicle in platoon $P_k$ to the RSU. Consider $V_{k,1}$ is driving towards east at $t_0$. Let $V_{k,i}$ be the target vehicle. During each time interval $\Delta t$, we assume that each vehicle follows the uniformly accelerated motion. Denote $v_{k,i}(t)$ as the velocity of $V_{k,i}$ at time $t$ and $a_{k,i}(t)$ as its acceleration rate. When $V_{k,i}$ is driving along the x-axis, $y_{k,i}(t)$ is fixed and $x_{k,i}(t)$ is calculated as

\begin{equation}
{x_{k,i}}(t+\Delta t)\! =\! {x_{k,i}}(t \!) \!+\! {v_{k,i}}(t\!)\Delta t + \frac{1}{2}{a_{k,i}}(t \!)\Delta {t^2}.
\label{eq4}
\end{equation}

When $V_{k,i}$ is driving along the y-axis, $x_{k,i}(t)$ is constant and $y_{k,i}(t)$ can be calculated as

\begin{equation}
{y_{k,i}}(t+\Delta t)\! =\! {y_{k,i}}(t \!) \!+\! {v_{k,i}}(t\!)\Delta t + \frac{1}{2}{a_{k,i}}(t \!)\Delta {t^2}.
\label{eq5}
\end{equation}

\begin{figure}
\center
\includegraphics[scale=0.45]{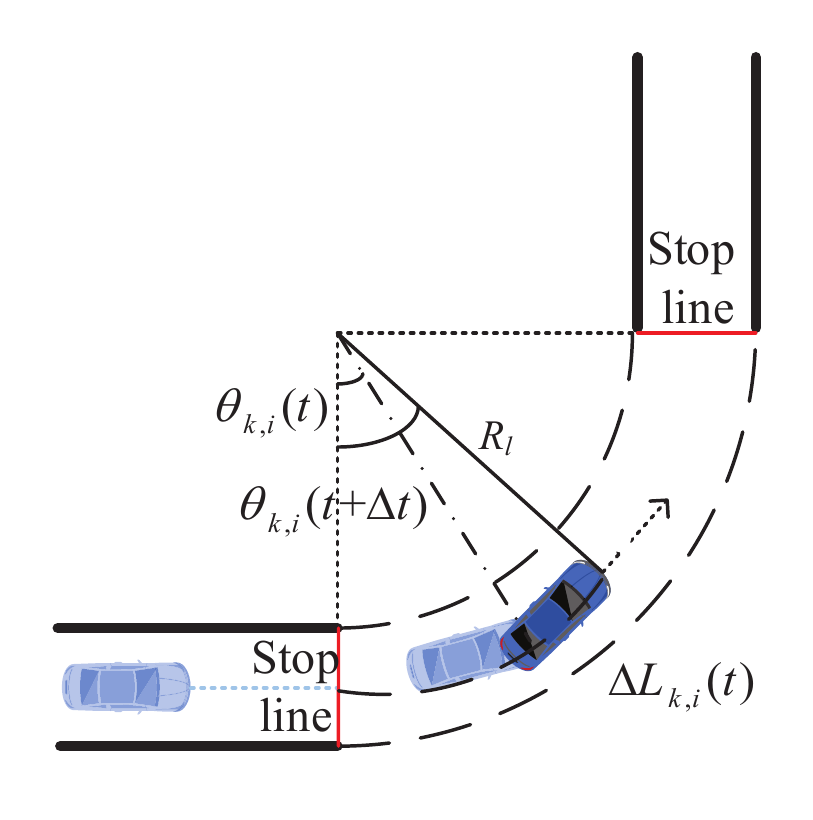}
\caption{Left-turn model.}
  \label{fig3}
\end{figure}

When $V_{k,i}$ is turning at the center area of the intersection, it would make a quarter circular motion after passing the stop lane until it arrives at the target lane. As shown in Fig. \ref{fig3}, circle center is the junction between the stop lane of the current lane and the target lane, and the radius is the distance from the center of stop line to the junction. Let $R_l$ and $R_r$ be the radius of the circle when $V_{k,i}$ is turning left and right, respectively, ${\theta _{k,i}}(t)$ be the angle between $V_{k,i}$'s driving direction and the direction of x-axis  at time $t$. Here, ${\theta _{k,i}}(t)$ is $0$ when $V_{k,i}$ begins to take a turn. We first consider the case when it is turning left. As shown in Fig. \ref{fig3}, the time-dependent angle of $V_{k,i}$ can be expressed as
\begin{equation}
{\theta _{k,i}}(t+\Delta t) = {\theta _{k,i}}(t) + \frac{\Delta L_{k,i}(t)}{{{R_l}}} + o(\Delta t) ,
\label{eq6}
\end{equation}where $\Delta L_{k,i}(t)$ is the driving distance of $V_{k,i}$ from time $t$ to $t+\Delta t$, which can be expressed as

\begin{equation}
\Delta L_{k,i}(t) =   {v_{k,i}}(t\!)\Delta t + \frac{1}{2}{a_{k,i}}(t \!)\Delta {t^2} + o(\Delta t^2) .
\label{eq7}
\end{equation}

Based on $\theta _{k,i}(t)$ and $\Delta L_{k,i}(t)$, the time-dependent position of $V_{k,i}$ is calculated as

\begin{equation}
\left\{ \begin{array}{l}
 {x_{k,i}}(t+\Delta t) \buildrel\textstyle.\over=  {x_{k,i}}(t) + \Delta L_{k,i}(t)\sin ({\theta _{k,i}}(t+\Delta t)) \\
 {y_{k,i}}(t+\Delta t) \buildrel\textstyle.\over=  {y_{k,i}}(t) + \Delta L_{k,i}(t)\cos ({\theta _{k,i}}(t+\Delta t)) \\
 \end{array} \right..
\label{eq8}
\end{equation}

Similarly, when turning right, the time-dependent position of $V_{k,i}$ can be calculated according to Eq. \eqref{eq8} by replacing $R_l$ with $R_r$ in Eq. \eqref{eq6}.

According to Eqs. \eqref{eq4}-\eqref{eq8}, we need to further derive $v_{k,i}(t)$ and ${a_{k,i}}(t \!)$ to determine the time-dependent position. Based on the uniformly accelerated motion equation, $v_{k,i}(t)$ is represented as

\begin{equation}
v_{k,i}(t+\Delta t) = v_{k,i}(t) + a_{k,i}(t)\Delta t + o(\Delta t) .\\
\label{eq9}
\end{equation}

In Eq. \eqref{eq9}, $v_{k,i}(t)$ is a function of ${a_{k,i}}(t)$. Therefore, we also need to derive ${a_{k,i}}(t)$ of leader vehicle and follower vehicles to determine the time-dependent position of $V_{k,i}$.

\subsubsection{Leader Vehicle}
\
\newline
\indent

If the inter-platoon spacing of leader vehicle $V_{k,1}$ towards its preceding vehicle is smaller than the safe intra-platoon spacing $s_e$, it would act as a follower vehicle to adjust its acceleration according to the IDM model, which will be introduced in the accelerations of follower vehicles. Otherwise, $V_{k,1}$ would select different accelerations when it reaches different areas, i.e., 1) enter the coverage of RSU; 2) enter the intersection area; 3) enter the center area of the intersection; 4) leave the center area of the intersection. According to the IDM model which is a car-following model widely adopted by platoons \cite{Jia2014}, $s_e$ is determined by $v_e$ and is calculated as

\begin{equation}
{s_e} = \frac{{{s_0} + {v_e}{T_0}}}{{\sqrt {1 - {{\left( {\frac{{{v_e}}}{{{v_0}}}} \right)}^4}} }}.
\label{eqs_e}
\end{equation}
where $v_0$ is the maximum velocity allowed on the road, $s_0$ is the minimum intra-platoon spacing and $T_0$ is the desired time headway. The time-dependent acceleration of leader vehicle $V_{k,1}$ is calculated as follows.
\paragraph{Enter the coverage of RSU}
\
\newline
\indent
 At the initial time $t_0$, $V_{k,1}$ enters the coverage of the RSU and sends the movement characteristics of each vehicle in $P_k$ to the RSU. Then $V_{k,1}$ keeps moving with constant velocity $v_e$ before it enters the intersection area, thus the time-dependent acceleration of $V_{k,1}$ is given as
\begin{equation}
{a_{k,1}}(t)=0,
\label{eq10}
\end{equation}
thus ${x_{k,1}}({t})$ is calculated according to Eq. \eqref{eq4} and ${y_{k,1}}({t})$ keeps ${y_{k,1}}({t_0})$.

\paragraph{Enter the intersection area}
\
\newline
\indent
When ${x_{k,1}}(t) = -(D+D_s)$, $V_{k,1}$ enters the intersection area. At this time, $V_{k,1}$ observes the state of the traffic light and adjusts its acceleration accordingly.



\textcircled{1}{red light}
\
\newline
\indent
When the traffic light is red, $V_{k,1}$ slows down in real time until stopping at the stop line. Specifically, at the beginning of each time interval $\Delta t$, $V_{k,1}$ adjusts its deceleration based on a uniformly accelerated motion until reaching the stop line, thus the speed gradually becomes zero to avoid emergency braking. The time-dependent acceleration of $V_{k,1}$ is calculated as

\begin{equation}
a_{k,1}(t) =  \frac{{v_{k,1}{{(t)}^2}}}{{2({D_s} + {x_{k,1}(t)})}}.
\label{eq12}
\end{equation}

\textcircled{2}{green light}
\
\newline
\indent
Let $t_1$ be the time when $V_{k,1}$ is at a location where ${x_{k,1}}(t) = -(D+D_s)$. When the traffic light is green, $V_{k,1}$ first observes the remaining time of green light $T_G^{rem}$ and estimates the position of $V_{k,1}$ at $t_1+T_G^{rem}$ when it drives with maximum acceleration $a$ specified in the IDM model. Since $V_{k,1}$ keeps driving with velocity $v_e$ before it enters the intersection area, $t_1$ is calculated as

\begin{equation}
t_{1} =  t_{0} - \frac{{x_{k,1}{(t_{0})}+(D+D_{s})}}{{v_{e}}}.
\label{eq11}
\end{equation}

Then, it will predict whether it can pass the stop line within $T_G^{rem}$. Note that if the velocity of $V_{k,1}$ exceeds the maximum velocity $v_0$, it would keep moving with the velocity $v_0$ \cite{Feng2019}. In other words, the acceleration of $V_{k,1}$ is

\begin{equation}
a _{k,1}(t) \!=\! \!\left\{\! \begin{array}{l}
 \begin{array}{*{20}{c}}
   a, & {v_{k,1}(t) < v_0}  \\
\end{array} \\
 \begin{array}{*{20}{c}}
   0, & {v_{k,1}(t) = v_0}  \\
\end{array} \\
 \end{array} \right..
\label{eq13}
\end{equation}

$V_{k,1}$ calculates ${x_{k,1}}({t_1+T_G^{rem}})$ according to Eqs. \eqref{eq4}-\eqref{eq9}. Note that the computation capacity of $V_{k,1}$ is powerful and it can calculate its position within a short time that can be neglected. Then $V_{k,1}$ takes different actions according to the prediction result. Specifically, if $ {x_{k,1}}({t_1+T_G^{rem}}) < -D_s$, $V_{k,1}$ cannot pass the stop line. In this case, it has to slow down in real time based on the acceleration in Eq. \eqref{eq12}. If ${x_{k,1}}({t_1+T_G^{rem}})\geq -D_s$, $V_{k,1}$ can pass the stop line by adjusting the acceleration based on Eq. \eqref{eq13}.

In this area, ${x_{k,1}}({t})$ is calculated according to Eq. \eqref{eq4} and ${y_{k,1}}({t})$ keeps ${y_{k,1}}({t_0})$.

\paragraph{Enter the center area of the intersection}
\
\newline
\indent
When $ x_{k,1}(t) = -D_s $, $V_{k,1}$ enters the center area of the intersection. At this time, it decelerates its velocity with the suitable deceleration $b$. Otherwise, it accelerates with acceleration $a$. In the deceleration or acceleration process, the velocity of $V_{k,1}$ would keep $v_e$ after it reaches $v_e$. Thus, the acceleration of $V_{k,1}$ is calculated as

\begin{equation}
a _{k,1}(t) \!=\! \!\left\{\! \begin{array}{l}
 \begin{array}{*{20}{c}}
   -b, & {v_{k,1}(t) > v_e}  \\
\end{array} \\
 \begin{array}{*{20}{c}}
   0, & {v_{k,1}(t) = v_e}  \\
\end{array} \\
 \begin{array}{*{20}{c}}
   a, & {v_{k,1}(t) < v_e}  \\
\end{array} \\
 \end{array} \right..
\label{eq14}
\end{equation}

In this area, $x_{k,1}(t)$ and $y_{k,1}(t)$ are calculated according to Eqs. \eqref{eq6}-\eqref{eq8} if $V_{k,1}$ is turning left. If $V_{k,1}$ is turning right, $x_{k,1}(t)$ and $y_{k,1}(t)$ can be calculated according to Eq. \eqref{eq8} by replacing $R_l$ with $R_r$ in Eq. \eqref{eq6}. If $V_{k,1}$ is driving straightly, ${x_{k,1}}({t})$ is calculated according to Eq. \eqref{eq4} and ${y_{k,1}}({t})$ keeps ${y_{k,1}}({t_0})$.

\paragraph{Leave the center area of the intersection}
\
\newline
\indent
When $\left| y_{k,1}(t)\right| = D_s$ or $x_{k,1}(t) = D_s$, $V_{k,1}$ leaves the center area of the intersection. At this time, $V_{k,1}$ would decelerate or accelerate with an acceleration calculated by Eq. \eqref{eq14}.

In this area, when $V_{k,1}$ turns left and right, ${y_{k,1}}({t})$ is calculated according to Eq. \eqref{eq5} and ${x_{k,1}}({t})$ keeps constant. When it drives straightly, ${x_{k,1}}({t})$ is calculated according to Eq. \eqref{eq4} and ${y_{k,1}}({t})$ keeps constant.

Note that the movement models are similar when $V_{k,1}$ drives toward other directions within the coverage of the RSU at $t_0$. In addition, the movement behaviors may be different at different intersections, but the movement model can be generically constructed in different cases if each vehicle follows the uniformly accelerated motion in each time interval $\Delta t$.
\subsubsection{Follower Vehicles}
\
\newline
\indent

Follower vehicles can adjust their accelerations to maintain the stable formation of platoons according to the IDM model.
The accelerations of the follower vehicles are calculated according to the IDM model \cite{Peng2017}, i.e.,
\begin{equation}
a _{k,i}(t) \!=a\left[ {1 - {{\left( {\frac{{v_{k,i}(t)}}{{{v_0}}}} \right)}^4} - {{\left( {\frac{{S_{k,i}^*(t)}}{{\Delta S_{k,i}(t)}}} \right)}^2}} \right],\\
\label{eq15}
\end{equation}where

\begin{equation}
S_{k,i}^*(t) = {s_0} + v_{k,i}(t){T_0} + \frac{{v_{k,i}(t)\Delta v_{k,i}(t)}}{{2\sqrt {ab} }}.
\label{eq16}
\end{equation}

In Eqs. \eqref{eq15} and \eqref{eq16}, $a$ is the maximum acceleration, $b$ is the suitable deceleration, $\Delta S_{k,i}(t)$ is the intra-platoon spacing between $V_{k,i}$ and $V_{k,i-1}$ at time $t$, $S_{k,i}^*(t)$ indicates the desired gap toward the preceding vehicle at time $t$ and $\Delta v_{k,i}(t)$ is the velocity difference between vehicle $V_{k,i}$ and the preceding vehicle $V_{k,i-1}$ at time $t$. In Eqs. \eqref{eq15} and \eqref{eq16}, $a$, $b$, $v_0$, $s_0$ and $T_0$ are given, and $\Delta v_{k,i}(t)$ is calculated as

\begin{equation}
\Delta v_{k, i}(t)=v_{k, i}(t)-v_{k, i-1}(t),
\label{eq17}
\end{equation}where $v_{k,i}(t)$ and $v_{k,i-1}(t)$ are calculated according to Eq. \eqref{eq9}. In addition,

\begin{small}
\begin{equation}
\Delta S_{k, i}(t)=\sqrt{\left[x_{k, i-1}(t)-x_{k, i}(t)\right]^{2}+\left[y_{k, i-1}(t)-y_{k, i}(t)\right]^{2}}-L_{0}.
\label{eq18}
\end{equation}
\end{small}where $L_0$ is the length of each vehicle, ${x_{k,i-1}}(t)$, ${x_{k,i}}(t)$, ${y_{k,i-1}}(t)$, ${y_{k,i}}(t)$ are calculated based on Eqs. \eqref{eq4}-\eqref{eq8}, which are dependent on the areas where $V_{k,i}$ and $V_{k,i-1}$ are located.


According to the above analysis, given the initial movement characteristics of each vehicle including position, velocity and acceleration of each vehicle at $t_0$, RSU can determine the time-dependent position of each vehicle in the intersection.






\subsection{Hearing Network}

\label{sec4a}
We denote $(x_{k,i}(t),y_{k,i}(t))$ as the position of $V_{k,i}$ at time $t$, $R_{c}$ as the communication range of each vehicle, and $M$ as the number of vehicles. Then, the hearing network matrix $H(t)$, which is an $M \times M$ square symmetric matrix, can be modelled as

\begin{small}
\begin{equation}
H(t) \!=\! \left[ {\begin{array}{*{20}{c}}
   {{h_{11,11}}\left( t\right)}  & \cdots & {{h_{11,ki}}\left( t\right)} & \cdots & {{h_{11,Pn_P}}\left( t\right)} \\
   \vdots & \vdots & \vdots &\vdots \\
   {{h_{ki,11}}\left( t\right)} & \cdots & {{h_{ki,ql}}\left( t\right)} & \cdots & {{h_{ki,Pn_P}}\left( t\right)}\\
   \vdots & \vdots & \vdots &\vdots \\
   {{h_{Pn_P,11}}\left( t\right)} & \cdots & {{h_{Pn_P,ki}}\left( t\right)} & \cdots & {{h_{Pn_P,Pn_P}}\left( t\right)} \\
\end{array}}\right],
\label{eq19}
\end{equation}
\end{small}where

\begin{small}
\begin{equation}
{h_{ki,ql}}( t ) \!=\! \!\left\{\! {\begin{array}{*{20}{c}}
   {\begin{array}{*{20}{c}}
   1,  {\sqrt {{{[{x_{k,i}}(t) \!-\! {x_{q,l}}(t)]}^2} \!+\! {{[{y_{k,i}}(t) \!-\! {y_{q,l}}(t)]}^2}}  \!\le\! {R_{c}}}  \\
\end{array}}  \\
   {\begin{array}{*{20}{c}}
   0,  {\sqrt {{{[{x_{k,i}}(t) \!-\! {x_{q,l}}(t)]}^2} \!+\! {{[{y_{k,i}}(t) \!-\! {y_{q,l}}(t)]}^2}}  \!>\! {R_{c}}}  \\
\end{array}}  \\
\end{array}} \right..
\label{eq20}
\end{equation}
\end{small}

In Eqs. \eqref{eq19} and \eqref{eq20}, if $h_{ki,ql}(t)=1$, the distance between $V_{k,i}$ and $V_{q,l}$ is smaller than the communication range $R_{c}$ at time $t$, i.e., $V_{k,i}$ and $V_{q,l}$ could communicate with each other at $t$; otherwise, they cannot communicate with each other at $t$. The value of $h_{ki,ql}(t)$ can be calculated according to the time-dependent positions of $V_{k,i}$ and $V_{q,l}$, which has been obtained in the previous subsection, thus the hearing network can be determined to reflect the time-varying connectivity among vehicles.


\section{Modeling of 802.11p}
\label{sec5}
In this section, we construct models of 802.11p for platooning communications at the intersection based on the hearing network. We first adopt the PSFFA to model the dynamic behavior of a transmission queue, then consider the continuous backoff freezing and adopt the z-domain linear model to describe the access process of the 802.11p with four ACs. Finally, based on the time-dependent model of 802.11p, the packet transmission delay and packet delivery ratio are derived.

\subsection{Dynamic Behavior of Transmission Queue}
 We adopt the single-server first-come first-service (FCFS) queuing system with infinite capacity for each transmission queue of $V_{k,i}$. Packets arrive at the transmission queue $AC_m$ according to a poisson process with the arrival rate $\lambda _{m}$, which is a constant. Let $N_{k,i,m}(t)$ be the average number of packet in transmission queue $AC_m$ of $V_{k,i}$ at time $t$ and $\dot{N}_{k,i,m}(t)$ be the average change rate of $N_{k,i,m}(t)$, $\mu _{k,i,m}(t)$ be the service rate of $AC_m$ in vehicle $V_{k,i}$ and ${\rho _{k,i,m}}(t)$ be the server utilization, i.e., the probability of non-empty transmission queue of $AC_m$ in $V_{k,i}$. Thus $\dot{N}_{k,i,m}(t)$ can be calculated according to the fluid flow (FF) model, i.e.,

\begin{equation}
\dot{N}_{k,i,m}(t) =  - \mu _{k,i,m}(t)\rho _{k,i,m}(t) + \lambda _{m}.
\label{eq21}
\end{equation}

Similar to \cite{Xu2016}, the access process of $AC_m$ obeys a general distribution and the service time at each time is independent and identically distributed. Let $T_{k,i,m}(t)$ and $\sigma _{k,i,m}^2(t)$ be the mean and variance of the service time in $AC_m$ of $V_{k,i}$ at time $t$, and thus we have $T_{k,i,m}(t) =\frac{1}{\mu _{k,i,m}(t)}$. In this case, the transmission queue of $AC_m$ is regraded as an M/G/1 queue model, and the relationship between $N_{k,i,m}(t)$ and $\rho _{k,i,m}(t)$ can be obtained through pollaczek-khinchine (P-K) formula \cite{Meighem2006}, i.e.,
\begin{equation}
{N_{k,i,m}}(t) = {\rho _{k,i,m}}(t) + \frac{{{\rho _{k,i,m}^2(t)}(1 + c_{k,i,m}^2(t))}}{{2(1 - {\rho _{k,i,m}}(t))}},
\label{eq22}
\end{equation}
where $c_{k,i,m}^2(t)$ is the squared coefficient variation of service time in $AC_m$ of $V_{i,j}$ at time $t$. Based on Eq. \eqref{eq22}, $\rho _{k,i,m}(t)$ is calculated as

\begin{equation}
\begin{split}
&{\rho _{k,i,m}}(t) \\
& = \frac{{{N_{k,i,m}}(t) + 1 - \sqrt {N_{k,i,m}^2(t) + 2c_{k,i,m}^2(t){N_{k,i,m}}(t) + 1} }}{{1 - c_{k,i,m}^2(t)}}.
\end{split}
\label{eq23}
\end{equation}

Substituting Eq. \eqref{eq23} into \eqref{eq21}, $\dot{N}_{k,i,m}(t)$ can be derived according to PSFFA \cite{Xu2016}, where the PSFFA equation for $AC_m$ is

\begin{equation}
\begin{split}
&\dot{N}_{k,i,m}(t) \\
&= - \frac{{N_{k,i,m}(t) + 1 - \sqrt {N_{k,i,m}^2(t) + 2c_{k,i,m}^2(t)N_{k,i,m}(t) + 1} }}{{1 - c_{k,i,m}^2(t)}}\\
& \times \mu _{k,i,m}(t) + \lambda _{m},
\end{split}
\label{eq24}
\end{equation}

In Eq. \eqref{eq24}, the arrival rate $\lambda _m$ $(m=0,1,2,3)$ is a given constant at time $t$. Given $N_{k,i,m}(t_0)$, ${\dot{N}_{k,i,m}(t)}$ can be determined by adopting the runge-kutta algorithm \cite{Ji2018} to solve Eq. \eqref{eq24} in each time interval $\Delta t$ if both $c_{k,i,m}(t)$ and $\mu _{k,i,m}(t)$ are determined. Here, ${c_{k,i,m}^2(t)}$ is defined as \cite{Xu2016}:

\begin{equation}
c_{k,i,m}^2(t) = {\sigma _{k,i,m}^2(t)}{\mu _{k,i,m}^2(t)},
\label{eq25}
\end{equation}and the relationship between $T_{k,i,m}(t)$ and $\mu _{k,i,m}(t)$ is

\begin{equation}
\mu _{k,i,m}(t) =\frac{1}{T_{k,i,m}(t)}.
\label{eq26}
\end{equation}

Thus, the mean and variance of the service time, i.e., $T_{k,i,m}(t)$ and $\sigma _{k,i,m}^2(t)$, will be required to determine $c_{k,i,m}(t)$ and $\mu _{k,i,m}(t)$. Next we will further model the access process of the 802.11p EDCA mechanism to derive $T_{k,i,m}(t)$ and $\sigma _{k,i,m}^2(t)$.
\subsection{Modeling of the 802.11p Access Process}
\label{sec5-b}
The access process of the 802.11p (i.e., the 802.11p EDCA mechanism) consists of transmission and backoff processes, thus the delay of vehicle $V_{k,i}$'s $AC_m$ is an aggregated time duration of the transmission and backoff processes. We will consider the continuous backoff freezing and adopt the z-domain linear model proposed in \cite{YuanYao2013} to model the access process of the 802.11p.

Let $P_{{\rm{st}}}^{k,i,m}(z)$ be the probability generating function (PGF) of the service time of $V_{k,i}$'s $AC_m$, ${T_{tr}}(z)$ be the PGF of the transmission time, ${B_j^{k,i,m}}(z)$ be the PGF of the backoff time of $AC_m$ when the number of retransmissions is $j$, and $p_v^{k,i,m}(t)$ be the internal collision probability of $AC_m$ in $V_{k,i}$, i.e., the probability that there are other ACs of $V_{k,i}$ with higher priority transmitting at the same time. Based on the z-domain linear model, we have

\begin{small}
\begin{equation}
\!\left\{\! \begin{array}{l}
 P_{{\rm{st}}}^{k,i,m}(z)\! =\! {B_0^{k,i,m}}(z){T_{tr}}(z) , m=0\\
 P_{{\rm{st}}}^{k,i,m}(z)\! =\! (1 \!-\! {p_v^{k,i,m}}(t)){T_{tr}}(z)\!\sum\limits_{n = 0}^{{M_m^l}} {\!\left[ {{{(p_v^{k,i,m}(t))}^n}\!\prod\limits_{\mathclap{j = 0}}^n {B_j^{k,i,n}(z)} }\! \right]} \\
 + {(p_v^{k,i,m}(t))^{M_m^l + 1}}\!\prod\limits_{j = 0}^{{M_m^l}} {B_j^{k,i,m}(z)} , m=1,2,3  \qquad \\
 \end{array} , \right.
\label{eq27}
\end{equation}
\end{small}

Next, ${T_{tr}}(z)$ and ${B_j^{k,i,m}}(z)$ are calculated as follows. Assuming the size of each packet in each AC is the same, the transmission time $T_{tr}$ is calculated as

\begin{equation}
{T_{tr}} = \frac{{PH{Y_H}}}{{{R_b}}} + \frac{{MA{C_H} + E[P]}}{{{R_d}}} + \gamma ,
\label{eq28}
\end{equation}where $PHY_{H}$ is the header length at physical layer, $MAC_{H}$ is the header length at MAC layer, $R_b$ is the basic rate, $R_ d$ is the data rate, $\gamma$ is the propagation delay and $E[P]$ is the packet size. In Eq. \eqref{eq28}, ${{PH{Y_H}}}$, $MA{C_H}$, $R_b$, $R_d$, $\gamma$ and $E[P]$ are predefined, thus  ${T_{tr}}(z)$ can be expressed as

\begin{equation}
{T_{tr}}(z) = z^{T_{tr}} ,
\label{eq29}
\end{equation}

In addition, ${B_j^{k,i,m}}(z)$ can be expressed based on the z-domain linear model, i.e.,

\begin{small}
\begin{equation}
\begin{array}{l}
 {B_j^{k,i,m}}(z) =  \\
 \left\{ \begin{array}{l}
 \frac{1}{{{W_{0,0}}}}\sum\limits_{n = 0}^{\mathclap{{W_{0,0}} - 1}} {{{\left[ {{H_{k,i,m}}(z)} \right]}^n},m = 0}  \\
 \frac{1}{{{W_{m,j}}}}\sum\limits_{n = 0}^{\mathclap{{W_{m,j}} - 1}} {{{\left[ {{H_{k,i,m}}(z)} \right]}^n},m = 1,2,3,j \in [0,{M_m}{\rm{ - 1}}]}  \\
 \frac{1}{{{W_{m,{M_m}}}}}\sum\limits_{n = 0}^{\mathclap{{W_{m,{M_m}}} - 1}} {{{\left[ {{H_{k,i,m}}(z)} \right]}^n},m = 1,2,3,j \in [M_m,M_m^l]}  \\
 \end{array} \right. \\
 \end{array},
\label{eq30}
\end{equation}
\end{small}where ${H_{k,i,m}}(z)$ is the PGF of the average time that the backoff counter in $AC_m$ of $V_{k,i}$ takes to decrease by one.

If the channel is sensed busy, i.e., it is occupied by other vehicles or other ACs of $V_{k,i}$ with higher priority, the backoff counter would be frozen for $T_{tr}+AIFS_m$. Thus the PGF of the backoff freezing time ${F_m}(z)$ is expressed as

\begin{equation}
{F_m}(z) = {z^{{T_{tr}} + AIF{S_m}}}.
\label{eq31}
\end{equation}

Let $p_b^{k,i,m}(t)$ be the probability that the channel is sensed busy. According to the z-domain linear model in \cite{Wu2019}, when the continuous backoff freezing is considered, ${H_{k,i,m}}(z)$ can be calculated by the mason formula, i.e.,

\begin{equation}
\begin{array}{l}
 {H_{k,i,m}}(z) = \frac{1}{{1 - p_b^{k,i,m}(t){F_m}(z)}} \\
  \times \{ (1 - p_b^{k,i,m}(t)){z^\delta }[1 - p_b^{k,i,m}(t){F_m}(z)] \\
  + p_b^{k,i,m}(t){F_m}(z)(1 - p_b^{k,i,m}(t)){z^\delta }\}  \\
  = \frac{{(1 - p_b^{k,i,m}(t)){z^\delta }}}{{1 - p_b^{k,i,m}(t){F_m}(z)}} \\
 \end{array},
\label{eq32}
\end{equation}

Combining the above equations, we can find that $P_{{\rm{st}}}^{k,i,m}(z)$ depends on $p_b^{k,i,m}(t)$ and $p_v^{k,i,m}(t)$. Thus, we will further derive $p_b^{k,i,m}(t)$ and $p_v^{k,i,m}(t)$. Let ${w_{k,i,m}}(t)$ be the internal transmission probability of $AC_m$, i.e., the probability that the backoff counter of $AC_m$ decreases to zero. Since $p_v^{k,i,m}(t)$ is the probability that there are other ACs of $V_{k,i}$ with higher priorities transmitting packets, $p_v^{k,i,m}(t)$ is expressed as

\begin{equation}
p_v^{k,i,m}(t)\! =\! \!\left\{\! \begin{array}{l}
 \begin{array}{*{20}{c}}
   0 & {m = 0}  \\
\end{array} \\
 \begin{array}{*{20}{c}}
   {1 \!-\! \prod\limits_{n = 0}^{m - 1} {(1 \!-\! {w_{k,i,n}}(t))} } & {m \!=\! 1,2,3}  \\
\end{array} \\
 \end{array} \right..
\label{eq33}
\end{equation}

Since $p_b^{k,i,m}(t)$ is the probability that other vehicles in the communication range of $V_{k,i}$ or other ACs of $V_{k,i}$ are transmitting, $p_b^{k,i,m}(t)$ is calculated as

\begin{equation}
p_b^{k,i,m}(t) \!=\! 1 \!-\! {\left[ {{{(1 \!-\! {\tau _{k,i}}(t))}^{N_{c}^{k,i}(t) \!-\! 1}}\prod\limits_{\mathclap{\scriptstyle n = 0, \hfill \atop
  \scriptstyle n \ne m \hfill}}^3 {(1 \!-\! {w_{k,i,n}}(t))} } \right]^{{A_m} \!+\! 1}},
\label{eq34}
\end{equation}where $N_{c}^{k,i}(t)$ is the number of vehicles in the communication range of $V_{k,i}$ and can be determined by the hearing network $H(t)$, which has been obtained in Section~\ref{sec4}, i.e.,

\begin{equation}
N_{c}^{k,i}(t) = \sum\limits_{q = 1}^P {\sum\limits_{l = 1}^{{n_k}} {{h_{ki,ql}}(t)} }.
\label{eq35}
\end{equation}

For $AC_m$, $A_m$ more slots needs to be detected than $AC_0$, thus $A_m$ is calculated as
\begin{equation}
A_m = AIFSN_m - AIFSN_0.
\label{eq36}
\end{equation}

Meanwhile, $\tau _{k,i}(t)$ is calculated as
\begin{small}
\begin{equation}
\tau _{k,i}(t) = \sum\limits_{m = 0}^3 {{\tau _{k,i,m}}(t)},
\label{eq37}
\end{equation}
\end{small}where ${\tau _{k,i,m}}(t)$ is the external transmission probability of $AC_m$ for $V_{k,i}$, i.e., the probability that only the current $AC_m$ of $V_{k,i}$ is transmitting while $ACs$ with higher priorities of $V_{k,i}$ are not transmitting. Thus ${\tau _{k,i,m}}(t)$ can be calculated as

\begin{small}
\begin{equation}
\tau _{k,i,m}(t) \!=\! \!\left\{\! \begin{array}{l}
 \begin{array}{*{20}{c}}
   w_{k,i,m}(t) & {m = 0}  \\
\end{array} \\
 \begin{array}{*{20}{c}}
   {{w_{k,i,m}} \prod\limits_{n = 0}^{m - 1} {(1 \!-\! {w_{k,i,n}}(t))} } & {m \!=\! 1,2,3}  \\
\end{array} \\
 \end{array} \right..
\label{eq38}
\end{equation}
\end{small}

According to Eqs. \eqref{eq33}-\eqref{eq38}, both $p_v^{k,i,m}(t)$ and $p_b^{k,i,m}(t)$ are determined by ${w_{k,i,m}}(t)$, which can be calculated as \cite{Yao2013}

\begin{small}
\begin{equation}
\!\left\{\! \begin{array}{l}
 {w_{\rm{k,i,m}}}(t) = {\left[ {\frac{{{W_{0,0}} + 1}}{{2\left( {1 - p_b^{k,i,m}(t)} \right)}} + \frac{{1 - {\rho _{k,i,m}}(t)}}{{p_a^{k,i,m}(t)}}} \right]^{ - 1}},m = 0 \\
 {w_{k,i,m}}(t) \!=\! \frac{{1 \!-\! {{(p_v^{k,i,m}(t))}^{M_m^l \!+\! 1}}}}{{1 \!-\! p_v^{k,i,m}(t)}} \!\left[\! {\frac{{1 \!-\! {{(p_v^{k,i,m}(t))}^{M_m^l \!+\! 1}}}}{{1 \!-\! p_v^{k,i,m}(t)}} \!+\! \frac{{{W_{m,0}} \!-\! 1}}{{2\left( {1 \!-\! p_b^{k,i,m}(t)} \right)}}} \right. \\
  \!+\! \frac{{{W_{m,0}}p_v^{k,i,m}(t)\left[ {1 \!-\! {{\left( {2p_v^{k,i,m}(t)} \right)}^{{M_m}}}} \right]}}{{\left( {1 \!-\! p_b^{k,i,m}(t)} \right)\left( {1 \!-\! 2p_v^{k,i,m}(t)} \right)}} \!+\! \frac{{1 \!-\! {\rho _{k,i,m}}(t)}}{{p_a^{k,i,m}(t)}}  \qquad  \qquad  \quad \qquad \ ,\\
 {\left. { \!+\! \frac{{{2^{{M_m} \!-\! 1}}{W_{m,0}}{{(p_v^{k,i,m}(t))}^{M_m^l \!+\! 1}}(1 \!-\! {{(p_v^{k,i,m}(t))}^{M_m^l \!-\! {M_m}}})}}{{\left( {1 \!-\! p_b^{k,i,m}(t)} \right)\left( {1 \!-\! p_v^{k,i,m}(t)} \right)}}} \!\right]\!^{ - 1}},m \!=\! 1,2,3 \\
 \end{array} \right.
\label{eq39}
\end{equation}
\end{small}where $p_a^{k,i,m}(t)$ is the probability that a packet arrives at $AC_m$. Since packets arrive at each AC according to the possion distribution with rate $\lambda _{m}(t)$, $p_a^{k,i,m}(t)$ is calculated as

\begin{equation}
p_a^{k,i,m}(t) = \sum\limits_{n = 1}^\infty  {\frac{{{{({\lambda _{m}}\delta )}^n}}}{{n!}}{e^{ - {\lambda _{m}}\delta }} = 1 - } {e^{ - {\lambda _{m}}\delta }}.
\label{eq40}
\end{equation}

Substituting Eq. \eqref{eq40} to Eq. \eqref{eq39}, it can be found that ${w_{k,i,m}}(t)$ is related to ${\rho _{k,i,m}}(t)$, thus $P_{{\rm{st}}}^{k,i,m}(z)$ is determined by ${\rho _{k,i,m}}(t)$. According to the property of the PGF approach, the $T_{k,i,m}(t)$ is calculated as

\begin{equation}
T_{k,i,m}(t) = {\left. {\frac{{dP_{st}^{k,i,m}(z)}}{{dz}}} \right|_{z = 1}},
\label{eq41}
\end{equation}and $\sigma _{k,i,m}^2(t)$ is calculated as
\begin{small}
\begin{equation}
\sigma _{k,i,m}^2(t) \!=\! {\left. {\!\left[\! {\frac{{{d^2}P_{st}^{k,i,m}(z)}}{{d{z^2}}} \!+\! \frac{{dP_{st}^{k,i,m}(z)}}{{dz}} \!-\! {{(\frac{{dP_{st}^{k,i,m}(z)}}{{dz}})}^2}} \!\right]\!} \right|_{z = 1}}.
\label{eq42}
\end{equation}
\end{small}

 As $T_{k,i,m}(t)$ and $\sigma _{k,i,m}^2(t)$ are related to ${\rho _{k,i,m}}(t)$, the calculation of $T_{k,i,m}(t)$ and $\sigma _{k,i,m}^2(t)$ is a non-linear complex solution process. Therefore, we use the iterative method to calculate $T_{k,i,m}(t)$ and $\sigma _{k,i,m}^2(t)$ step by step.
\begin{itemize}
\item[(1)] Firstly, initialize $\rho_{k,i,m}(t)$ in the range $(0,1)$;
\item[(2)] Secondly, calculate $p_b^{k,i,m}(t)$ and $p_v^{k,i,m}(t)$ according to Eqs. \eqref{eq33} and \eqref{eq34}.
\item[(3)] Thirdly, calculate the mean service time $T_{k,i,m}(t)$ according to Eq. \eqref{eq41} and calculate a new value of $\rho _{k,i,m}(t)$ according to the equation $min({\lambda _{m}}T_{k,i,m}(t), 1)$.
\item[(4)] Fourthly, compare the absolute error between the initial and new value of $\rho _{k,i,m}(t)$ with the predefined error bound $\varepsilon$. If the absolute error is less than $\varepsilon$, the iteration is terminated with the current $\rho _{k,i,m}(t)$ as an output. Otherwise, repeat the above iteration with the updated $\rho_{k,i,m}(t)$ as an initial value until the absolute error is less than $\varepsilon$.
\item[(5)] Finally, calculate the variance of service time $\sigma _{k,i,m}^2(t)$ according to Eq. \eqref{eq42}.
\end{itemize}

Now, we have obtained $\tau _{k,i}(t)$, $\rho _{k,i,m}(t)$, $T_{k,i,m}(t)$ and $\sigma _{k,i,m}^2(t)$ based on which $c_{k,i,m}(t)$ and $\mu _{k,i,m}(t)$ can be calculated according to Eqs. \eqref{eq25} and \eqref{eq26}. Then $\dot{N}_{k,i,m}(t)$ in Eq. \eqref{eq24} can be further determined based on the PSFFA method \cite{Xu2016}.

\subsection{PTD and PDR}

Since the packets transmitted by $V_{k,i}$ should be received successfully within a short time duration at the intersection, PTD and PDR are very important for platooning communications.

\subsubsection{Packet transmission delay}
\
\newline
\indent

The packet transmission delay of vehicle $V_{k,i}$'s $AC_m$ at time $t$ is defined as the time duration of a packet from entering the transmission queue $AC_m$ of $V_{k,i}$ to being received successfully or discarded. According to the little's law, the average number of packets in transmission queue $AC_m$ of $V_{k,i}$ at time $t$ can be expressed as

\begin{equation}
N_{k,i,m}(t) = \lambda _{m} \times T_{k,i,m}^s(t),
\label{eq43}
\end{equation}where $T_{k,i,m}^s(t)$ is the average residence time of the packet in $AC_m$ of $V_{k,i}$ at time $t$. The change rate of $T_{k,i,m}^s(t)$ can be expressed as

\begin{equation}
\mathop {{T_{k,i,m}^s}}\limits^.(t)  = \frac{{{{\mathop {N_{k,i,m}}\limits^. }}(t)}}{{{\lambda _{m}}}}.
\label{eq44}
\end{equation}

Let $PTD_{k,i,m}(t)$ be the packet transmission delay of $V_{k,i}$'s $AC_m$ at time $t$. The packet transmission delay at time $t+ \Delta t$ can be calculated as the summation of $PTD_{k,i,m}(t)$ and the change of average residence time of the packet in $AC_m$ within $[t ,t+ \Delta t]$, i.e., $\int_{t}^{t+\Delta t} {\mathop {{T_{k,i,m}^s}}\limits^.(t)}dt$. Thus according to Eq. \eqref{eq44}, $PTD_{k,i,m}(t+ \Delta t)$ is calculated as

\begin{small}
\begin{equation}
\begin{array}{l}
PTD_{k,i,m}(t+ \Delta t) = \\
\left\{ \begin{array}{l}
 \frac{{{N_{k,i,m}}({t_0})}}{{{\lambda _{m}}}},t = {t_0} \\
 PT{D_{k,i,m}}(t) + \int_{t}^{t+\Delta t} {\frac{\dot{N}_{k,i,m}(t)}{{{\lambda _{m}}}}dt,t > {t_0}}  \\
 \end{array} \right.,
 \end{array}
\label{eq45}
\end{equation}
\end{small}where ${{N_{k,i,m}}({t_0})}$ is given. Since $\dot{N}_{k,i,m}(t)$ has been obtained in Section~\ref{sec5}, we can determine the time-dependent packet transmission delay according to Eq. \eqref{eq45}.

\subsubsection{Packet delivery ratio}
\
\newline
\indent

The packet delivery ratio of $AC_m$ for $V_{k,i}$ at time $t$ (denoted as $PDR_{k,i,m}(t)$) is calculated as the ratio between the number of successfully received packets (denoted as $f_s^{k,i,m}(t)$) and the total number of packets arriving at $AC_m$ of $V_{k,i}$ at time $t$ (denoted as $f_a^{k,i,m}(t)$). Thus, $PDR_{k,i,m}(t)$ is expressed as

\begin{equation}
PDR_{k,i,m}(t) = \frac{f_{s}^{k,i,m}(t)}{f_a^{k,i,m}(t)}.
\label{eq46}
\end{equation}

Here, $f_a^{k,i,m}(t)$ and $f_s^{k,i,m}(t)$ are derived from the hearing network $H(t)$. In particular, $f_a^{k,i,m}(t)$ is calculated as

\begin{equation}
f_a^{k,i,m}(t)={{\sum\limits_{q = 1}^P {\sum\limits_{l = 1}^{n_k} {\lambda_{m}{h_{ki,ql}(t)}}}}}.
\label{eq47}
\end{equation}

Meanwhile, as the number of transmitted packets of $AC_m$ for $V_{k,i}$ at time $t$ is ${\mu_{k,i,m}}(t){\rho _{k,i,m}}(t)$, $f_s^{k,i,m}(t)$ is expressed as

\begin{small}
\begin{equation}
f_s^{k,i,m}(t) = {{\sum\limits_{q = 1}^P {\sum\limits_{l = 1}^{n_k} {{\mu_{k,i,m}}(t){\rho_{k,i,m}}(t){h_{ki,ql}}}(t) {P_{{ki},{ql}}^s}(t)} }},
\label{eq48}
\end{equation}
\end{small}where ${P_{{ki},{ql}}^s}(t)$ is the probability that $V_{q,l}$ successfully receives a packet transmitted from $V_{k,i}$. Since physical layer has a much smaller impact on packet transmission compared to MAC layer, we consider the ideal channel in this paper, similar to the related work \cite{Yao2013}, i.e., the packet loss is only incurred by the collisions at MAC layer. Therefore, the successful transmission occurs when both conditions are satisfied. One is that no other vehicles in the communication range of $V_{k,i}$ is transmitting at the same time, i.e., there is no collision caused by exposed vehicles. Another condition is that no other vehicle in the communication range of $V_{k,i}$ but not in the communication range of $V_{q,l}$ is transmitting at the same time, i.e., there is no collision caused by the hidden vehicles. Let $P_{ki,ql}^{exposed}(t)$ be the collision probability that is caused by exposed vehicles, and $P_{ki,ql}^{hidden}(t)$ be the collision probability that is caused by hidden vehicles. Thus, we have

\begin{equation}
P_{ki,ql}^s(t) =( {1\! -\! P_{ki,ql}^{exposed}(t)})( {1\! -\! P_{ki,ql}^{hidden}(t)} ).
\label{eq49}
\end{equation}

The collision probability caused by exposed vehicles is the probability that there is at least one vehicle in the communication range of $V_{k,i}$ transmitting at the same time. Therefore, $P_{ki,ql}^{exposed}(t)$ is calculated as

\begin{small}
\begin{equation}
\begin{array}{*{20}{c}}
   {P_{ki,ql}^{exposed}(t) \!=\! {h_{ki,ql}} \times \left[ {1 \!-\! \prod\limits_{r = 1}^P {\prod\limits_{s = 1}^{{n_k}} {{{\left[ {(1 \! -\! {\tau _{r,s}}(t))} \right]}^{{h_{ki,rs}}(t)}}} } } \right]}  \\
   {\forall k,q,r = 1,...,P;\forall i,l,s = 1,...,{n_k}.}  \\
\end{array},
\label{eq50}
\end{equation}
\end{small}

Denote $T_{tr}$ as the transmission delay. The collision probability caused by hidden vehicles is the probability that there is at least one vehicle outside the communication range of $V_{k,i}$ transmitting to $V_{q,l}$ within the time period $[t-T_{tr},t+T_{tr}]$. Therefore, $P_{ki,ql}^{hidden}(t)$ is calculated as

\begin{small}
\begin{equation}
\begin{array}{*{20}{c}}
   \begin{array}{l}
 P_{ki,ql}^{hidden}(t) = {h_{ki,ql}} \\
  \times \left[ {1 - \prod\limits_{r = 1}^P {\prod\limits_{s = 1}^{{n_k}} {{{\left[ {(1 - {\tau _{r,s}}(t))} \right]}^{\frac{{2{T_{tr}}}}{{\delta}}(1 - {h_{ki,ql}}(t)){h_{ql,rs}}(t)}}} } } \right] \\
 \end{array}  \\
   {\forall k,q,r = 1,...,P;\forall i,l,s = 1,...,{n_k}.}  \\
\end{array},
\label{eq51}
\end{equation}
\end{small}

Now, we have obtained the collision probability caused by exposed vehicle and hidden vehicle. Since ${\mu_{k,i,m}}(t)$, ${\rho _{k,i,m}}(t)$ and ${\tau _{r,s}}(t)$ have been obtained in Section \ref{sec5-b}, we can determine the time-dependent packet delivery ratio according to Eqs. \eqref{eq46}-\eqref{eq51}. Specifically, substituting Eqs. \eqref{eq50} and \eqref{eq51} into Eq. \eqref{eq49}, the successful receiving probability can be achieved. Furthermore, after substituting Eq. \eqref{eq49} into Eq. \eqref{eq48}, the time-dependent packet delivery ratio can be determined by substituting Eqs. \eqref{eq47} and \eqref{eq48} into Eq. \eqref{eq46}.

\section{Simulation and Analytical Results}
\label{sec6}

\begin{figure}
\centering
\includegraphics[scale=0.25]{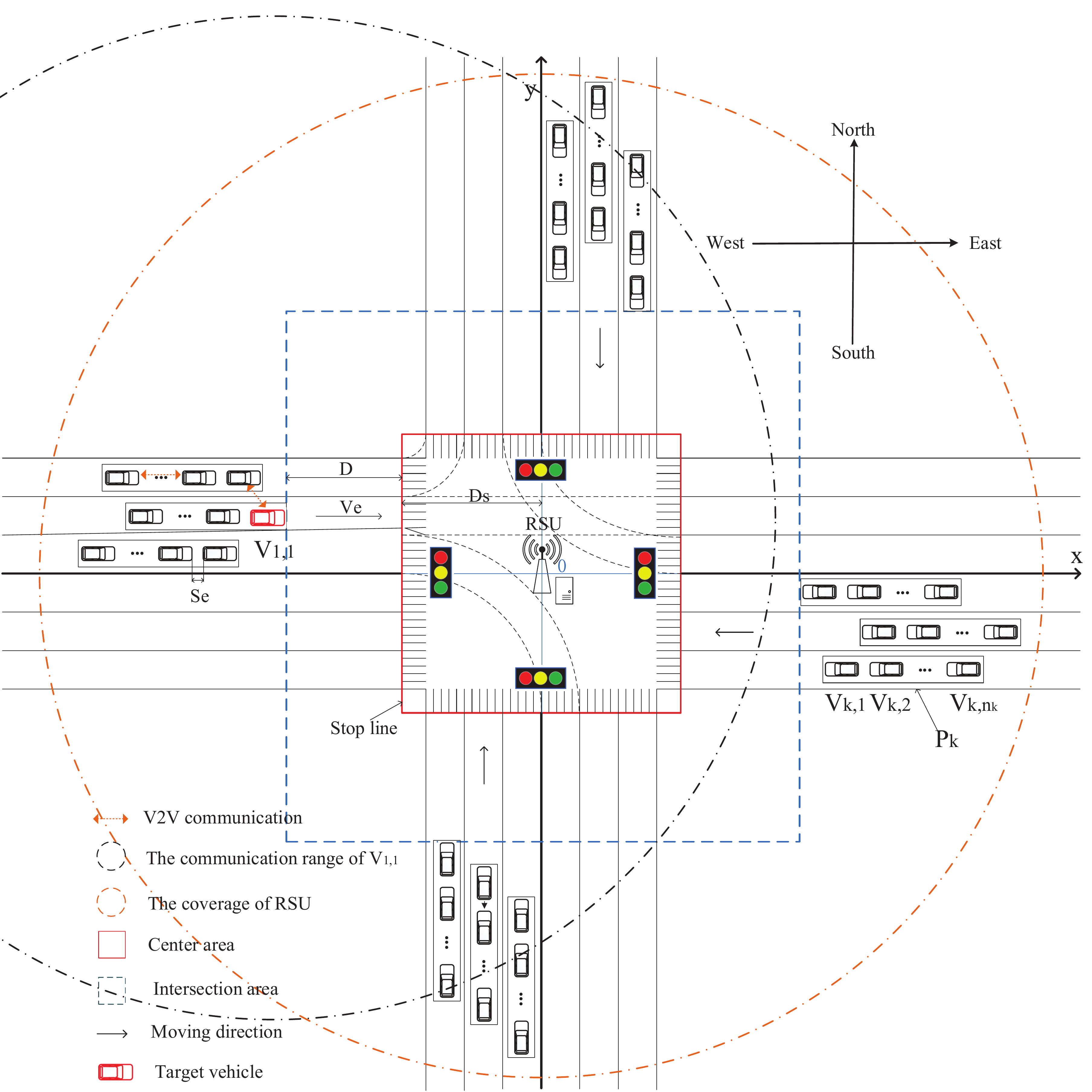}
\caption{Initial scenario.}
\label{simmodel}
\end{figure}

In this section, we present simulation experiments and compared the simulation results with the analytical results, i.e., the results obtained according to the analytically model, to verify the accuracy of the proposed analytical model. Furthermore, we evaluate the PTD and PDR of 802.11p for platooning communications at the intersection according to the analytical results. In order to investigate affection of other vehicles on a target vehicle during the whole movements at different intersection areas, we initially set that all the platoons are randomly distributed in the coverage of RSU and outside the intersection area. The target vehicle is $V_{1,1}$. The initial scenario is shown in Fig. \ref{simmodel}. The initial traffic light is green at the beginning. The related parameters used in both the simulation experiments and analytical model are listed in Table \ref{tab2}, where the parameters of 802.11p is set according the 802.11p standard \cite{Hafeez2016}. Note that we set the time interval $\Delta t$ as the maximum delay limit defined by ETSI to ensure that a packet can be accepted successfully, i.e., $100 ms$ \cite{European}.

\begin{table}[h]\tiny
\caption{RELATED PARAMETERS VALUES}
\label{tab2}
\centering
\begin{tabular}{cccc}
\toprule
\textbf{Parameter} &\textbf{Value} &\textbf{Parameter} &\textbf{Value}\\
\midrule
$a$ & $2m/{s^2}$ & $b$ & $3m/{s^2}$\\
\midrule
$AIFSN_{0}$ & $2$ & $AIFSN_{1}$ & $3$\\
\midrule
$AIFSN_{2}$ & $6$ & $AIFSN_{3}$ & $9$\\
\midrule
$CW_{min}^{0}$ & $3$ & $CW_{min}^{1}$ & $3$\\
\midrule
$CW_{min}^{2}$ & $7$ & $CW_{min}^{3}$ & $15$\\
\midrule
$CW_{max}^{0}$ & $3$ & $CW_{max}^{1}$ & $7$\\
\midrule
$CW_{max}^{2}$ & $15$ & $CW_{max}^{3}$ & $1023$\\
\midrule
$D$ & $4m$ & $D_{r}$ & $100m$\\
\midrule
$D_s$ & $16.5m$ &$\lambda_{0}$ & $5$ \\
\midrule
$\lambda_{1}$ & $10$ & $\lambda_{2}$ & $15$ \\
\midrule
$\lambda_{3}$ & $20$ & $v_{0}$ & $50mile/h$ \\
\midrule
$v_{e}$ & $25mile/h$  & $T_{0}$ & $1.5s/Veh$\\
\midrule
$s_{0}$ & $3m$ & $s_{e}$ & $4m$ \\
\midrule
$P$ & $24$ & $n_k$ & $3$\\
\midrule
$L_0$ & $3m$ & $R_c$ & $100m$\\
\midrule
$T_R$ & $150s$ & $T_G$ & $30s$\\
\midrule
$R_l$ & $7.75m$ & $R_r$ & $18.25m$\\
\midrule
$\Delta t$ & $0.1s$ & $\delta$  & $13 \mu s$\\
\midrule
$\gamma$  & $ 1 \mu s$ &  $E[P]$ &  $200bit$  \\
\midrule
$R_d$   &  $3Mbps$  &    $R_b$   &  $1Mbps$  \\
\midrule
$PH{Y_H}$ & $48bits$ & $MA{C_H}$ & $112bits$\\
\midrule
$SIFS$ & $32\mu s$   & Retransmission Limit ($L_m$)  & $1$   \\
\bottomrule
\end{tabular}
\end{table}

\begin{figure}
\centering
\includegraphics[scale=0.5]{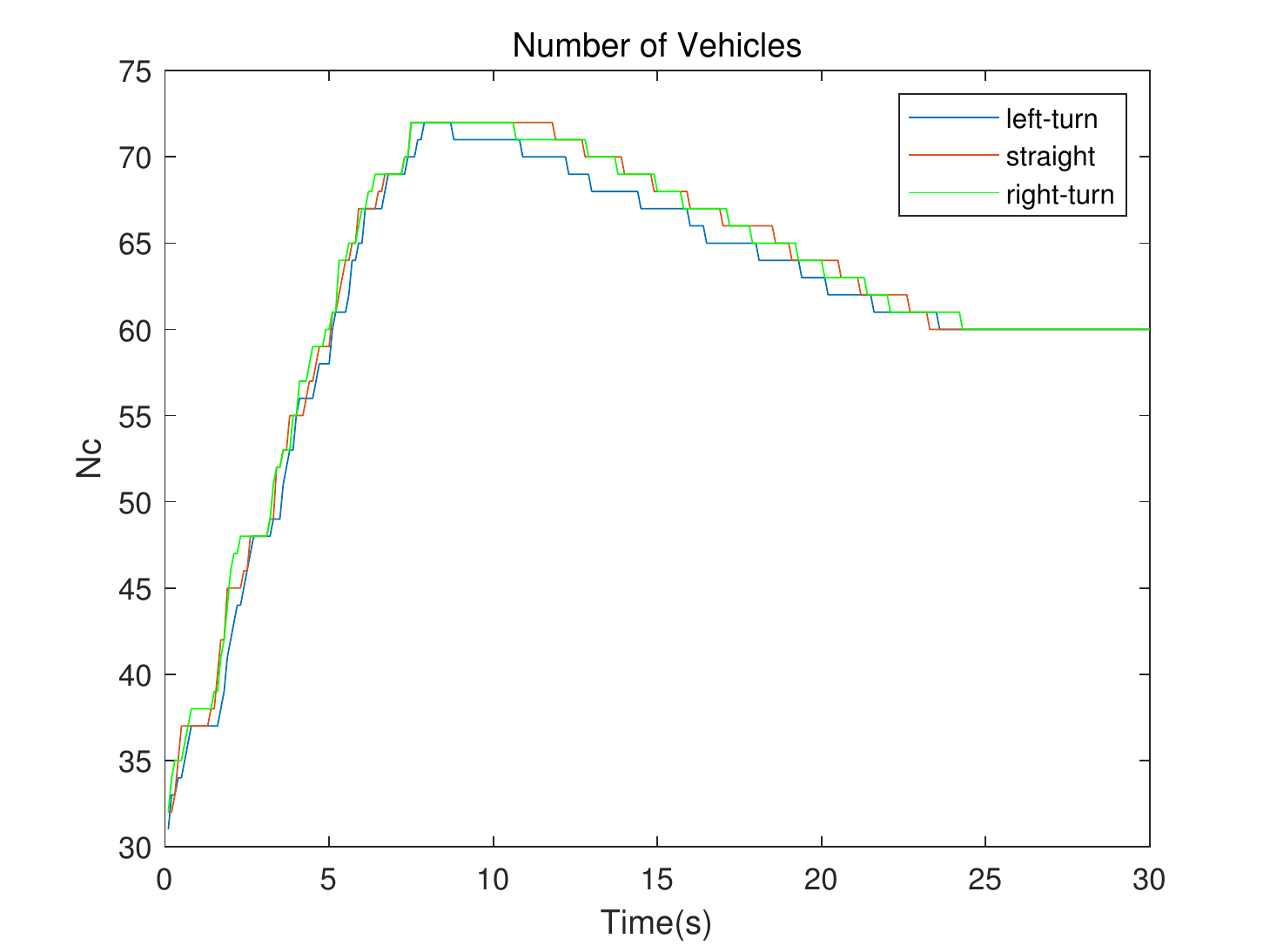}
\caption{ The time-dependent number of vehicles in the communication range of $V_{1,1}$.}
\label{fig4}
\end{figure}

\begin{figure*}
  \centering
  \subfigure[]{
    \begin{minipage}{5.7cm}
    \includegraphics[scale=0.44]{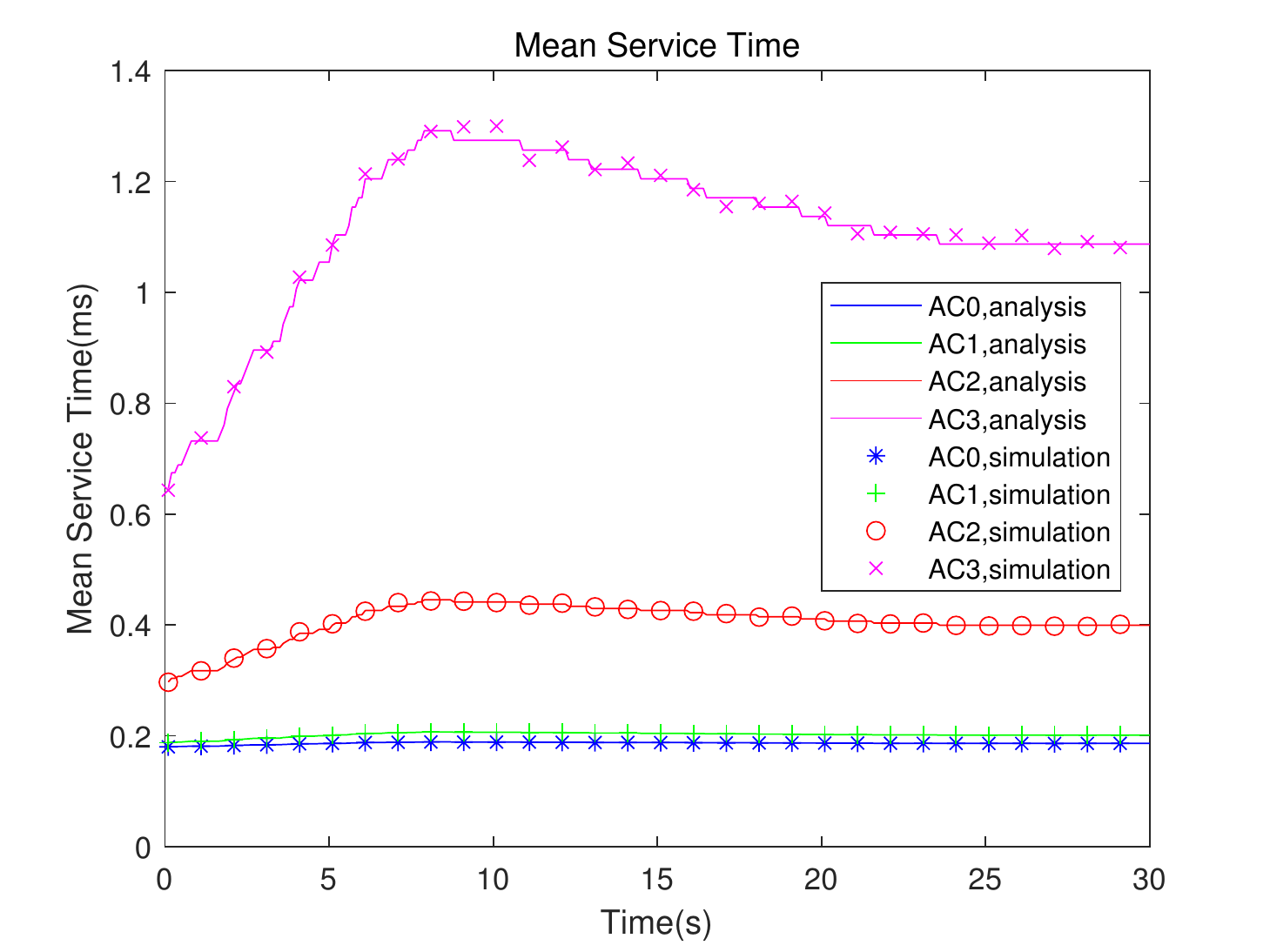}
    \end{minipage}}
  \subfigure[]{
    \begin{minipage}{5.7cm}
    \includegraphics[scale=0.44]{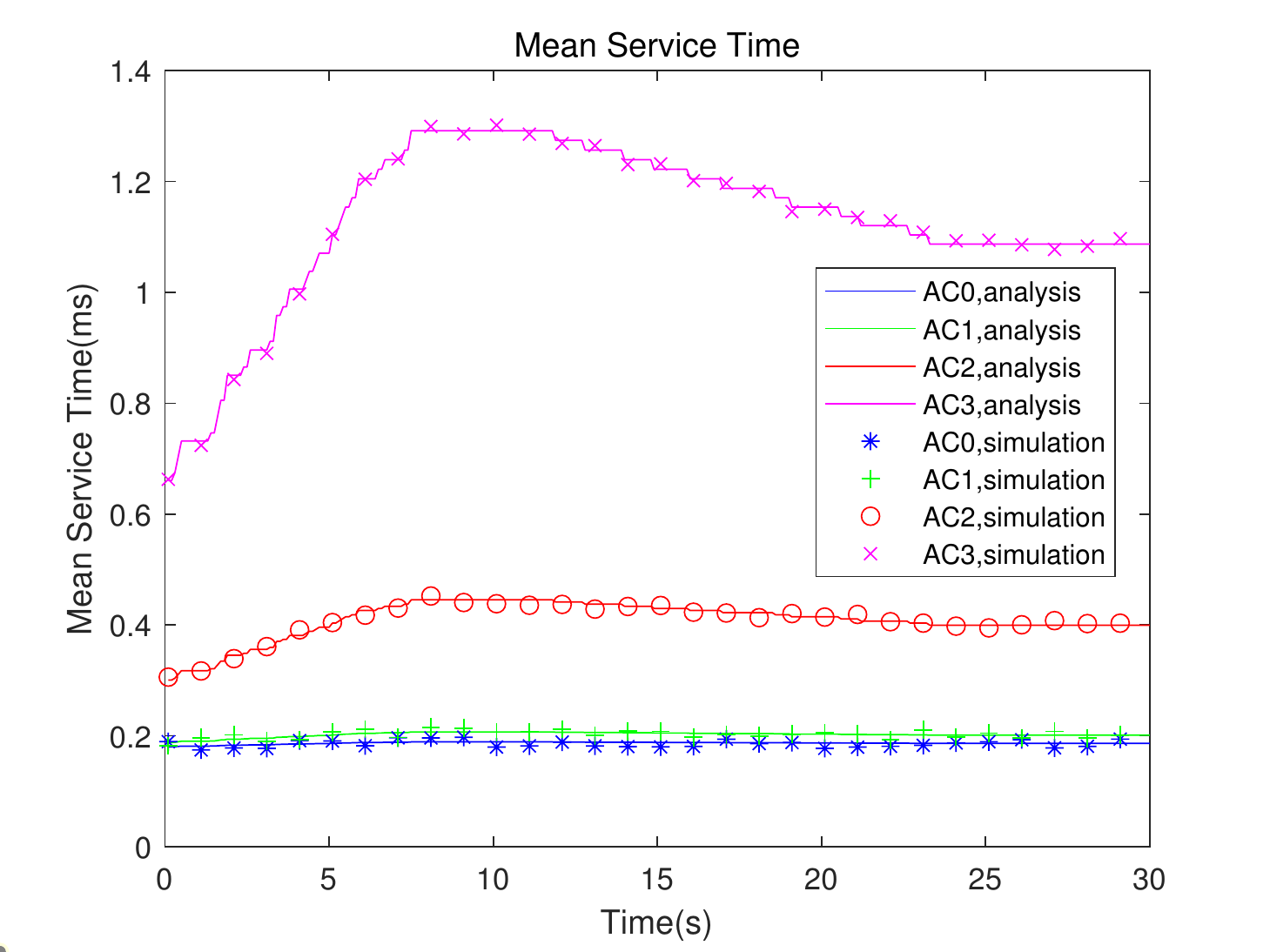}
    \end{minipage}}
  \subfigure[]{
    \begin{minipage}{5.7cm}
    \includegraphics[scale=0.44]{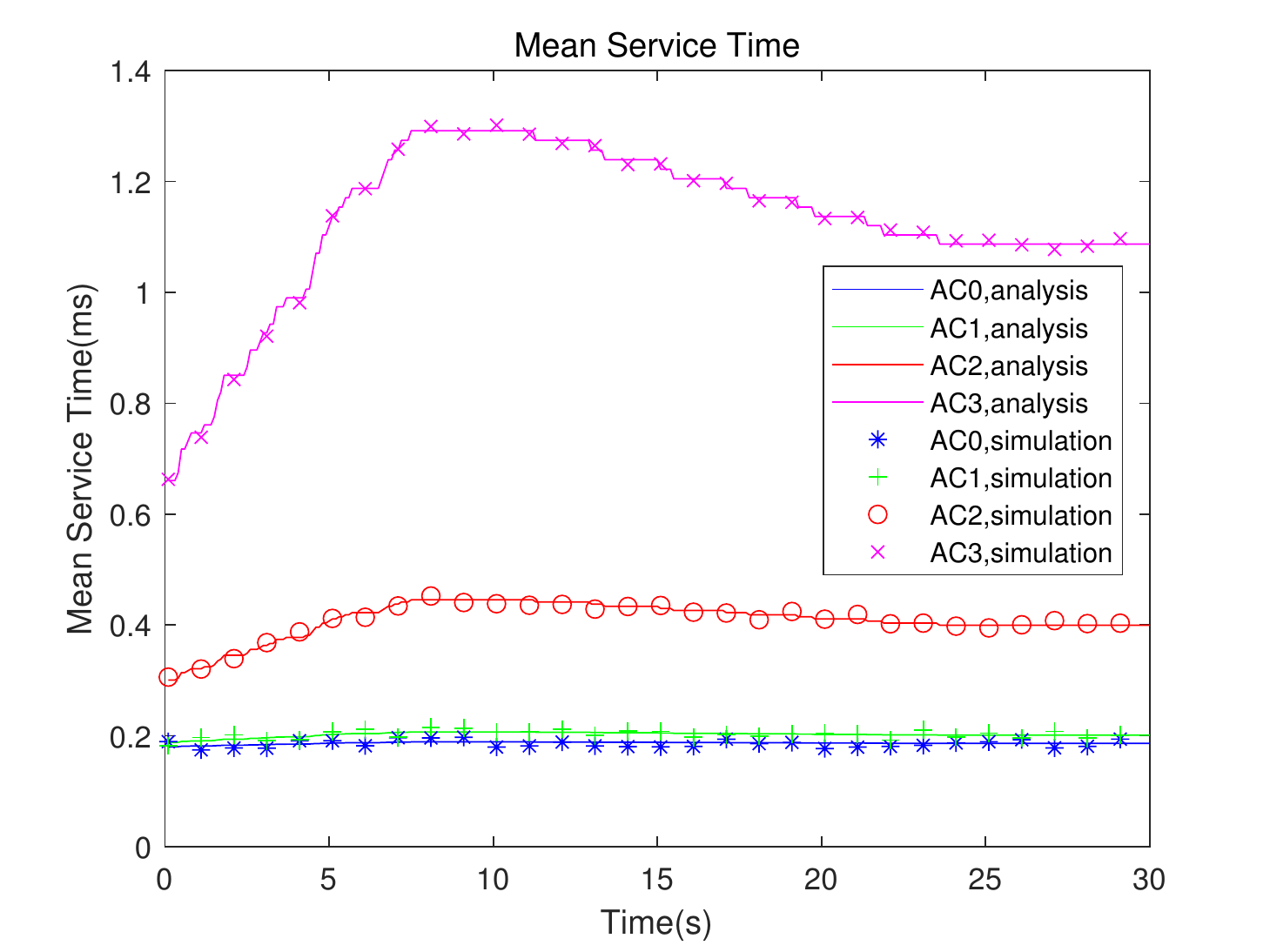}
    \end{minipage}}
    \caption{The time-dependent mean service time. (a) Left-turn; (b) Straight; (c) Right-turn.}
    \label{fig5}
    \vspace{-0.4cm}
\end{figure*}

\begin{figure*}
  \centering
  \subfigure[]{
    \begin{minipage}{5.7cm}
    \includegraphics[scale=0.44]{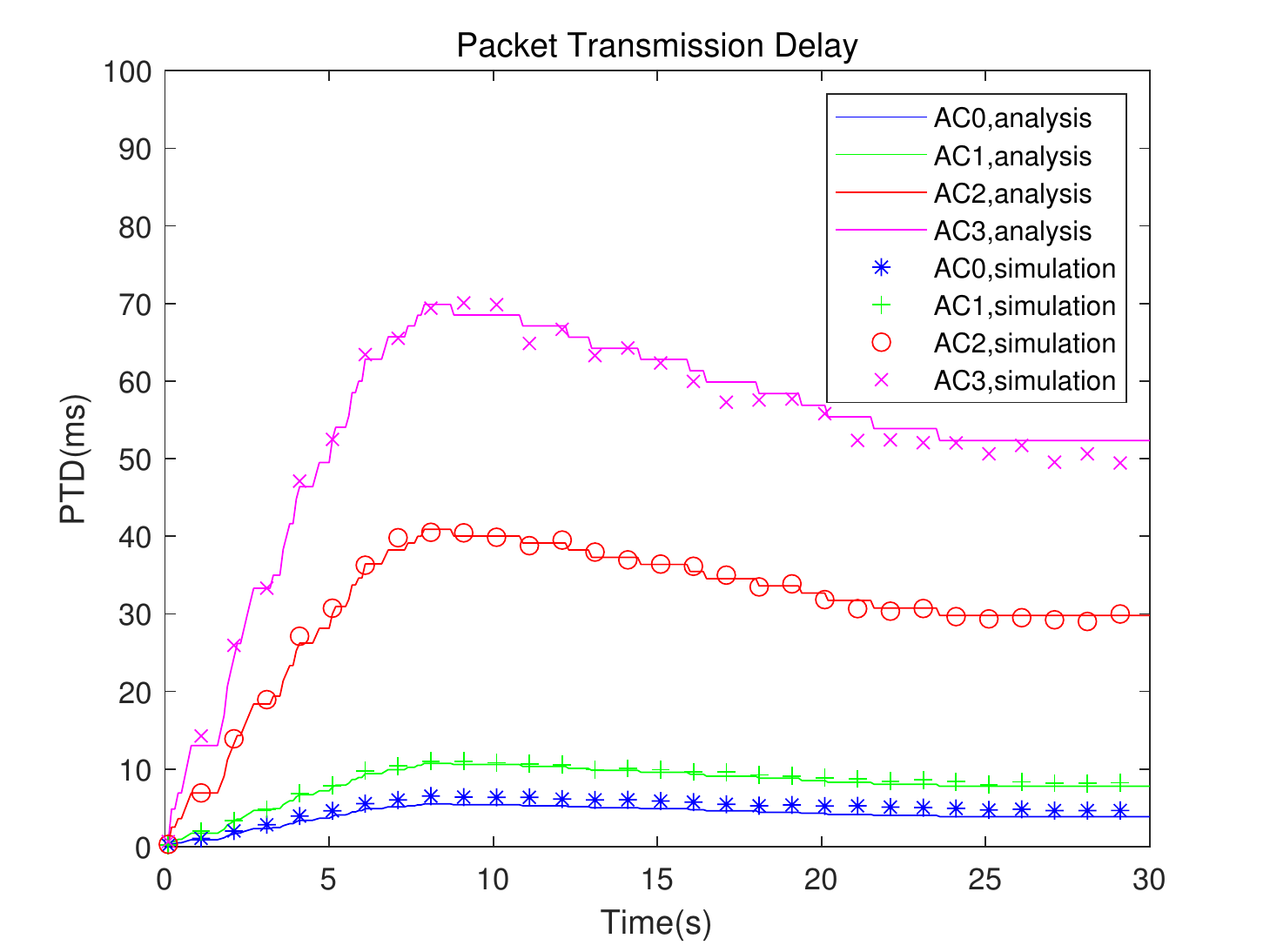}
    \end{minipage}}
  \subfigure[]{
    \begin{minipage}{5.7cm}
    \includegraphics[scale=0.44]{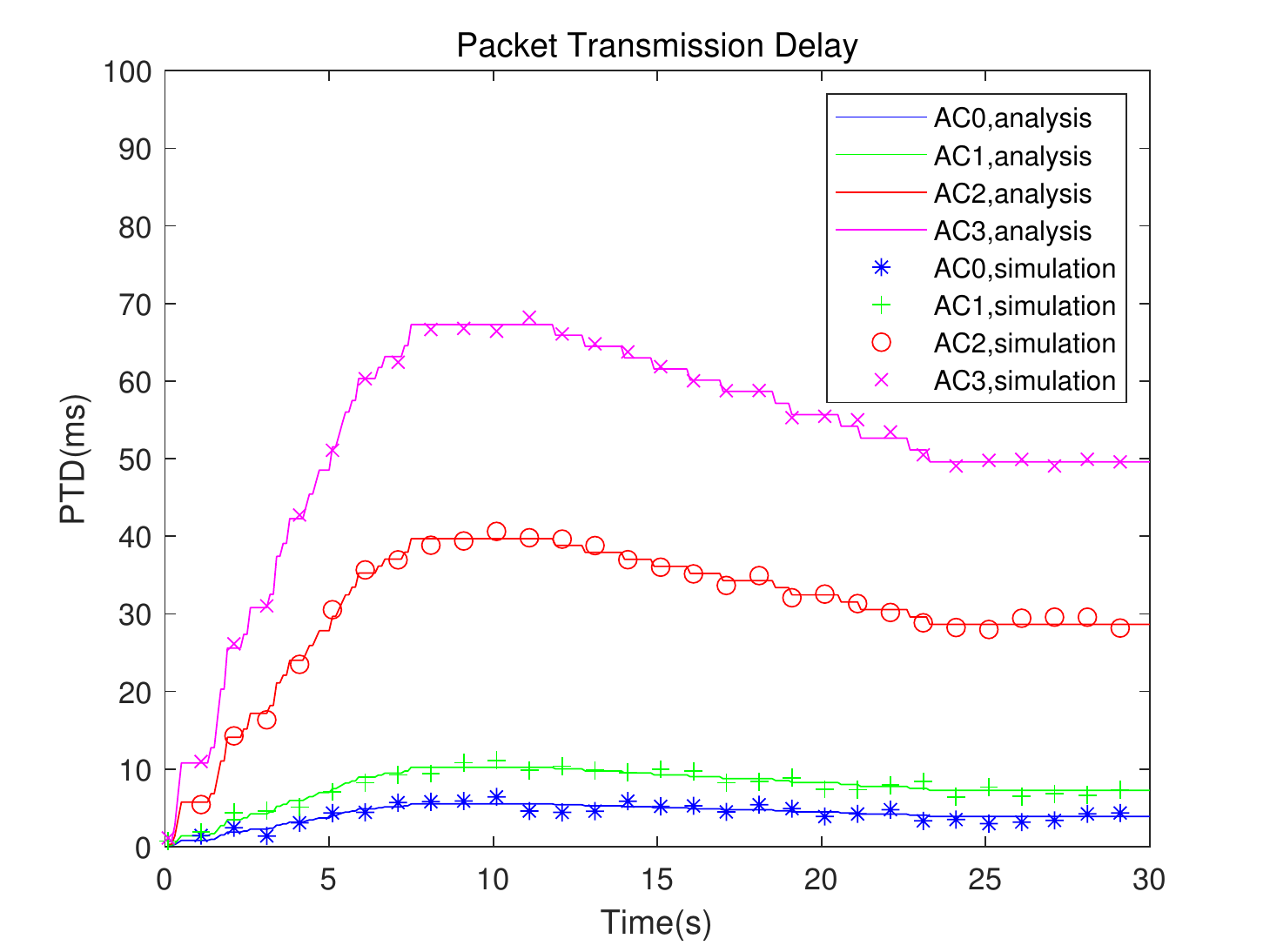}
    \end{minipage}}
  \subfigure[]{
    \begin{minipage}{5.7cm}
    \includegraphics[scale=0.44]{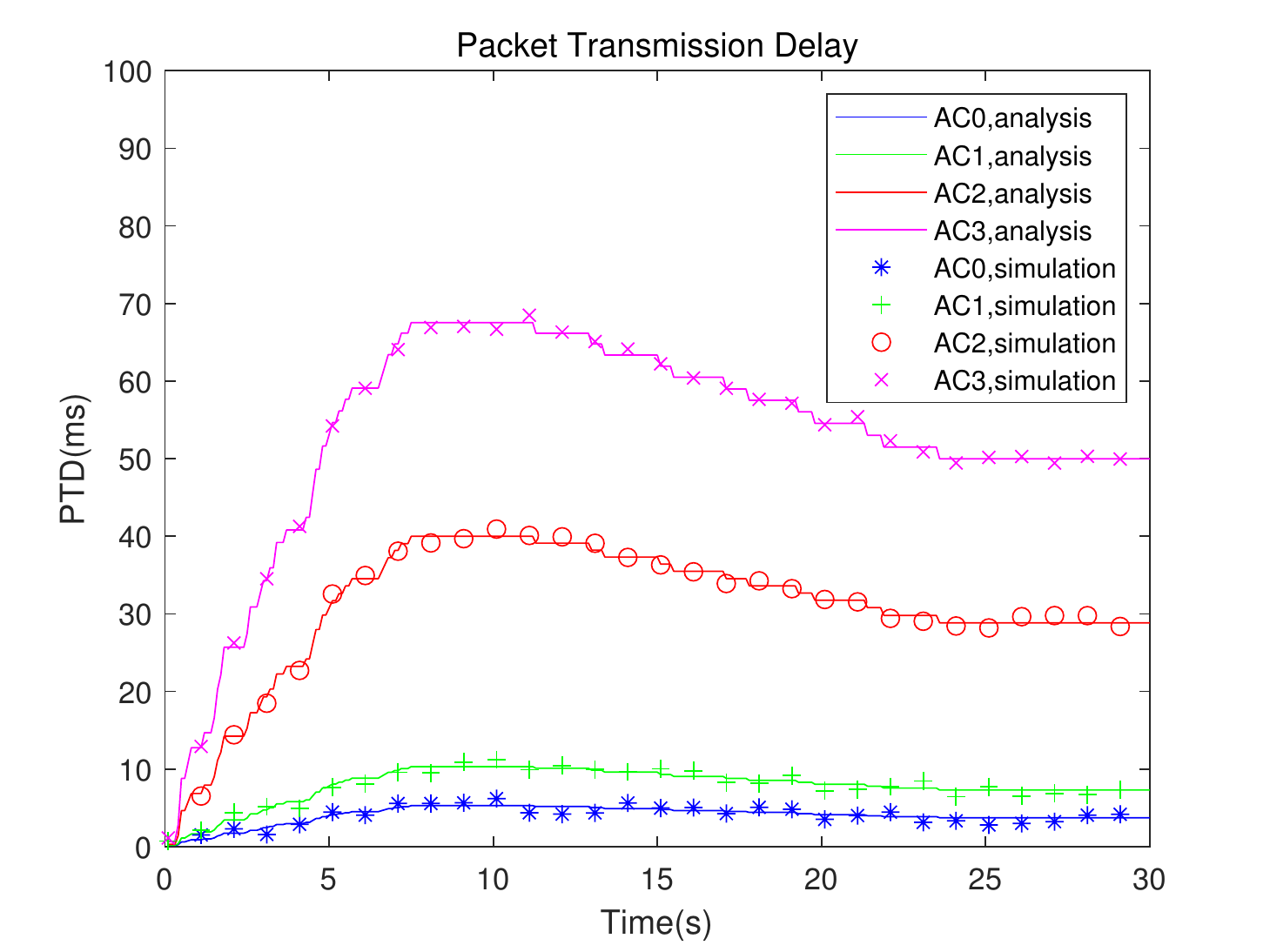}
    \end{minipage}}
    \caption{The time-dependent packet transmission delay. (a) Left-turn; (b) Straight; (c) Right-turn.}
    \label{fig6}
    \vspace{-0.4cm}
\end{figure*}

\begin{figure*}
  \centering
  \subfigure[]{
    \begin{minipage}{5.7cm}
    \includegraphics[scale=0.44]{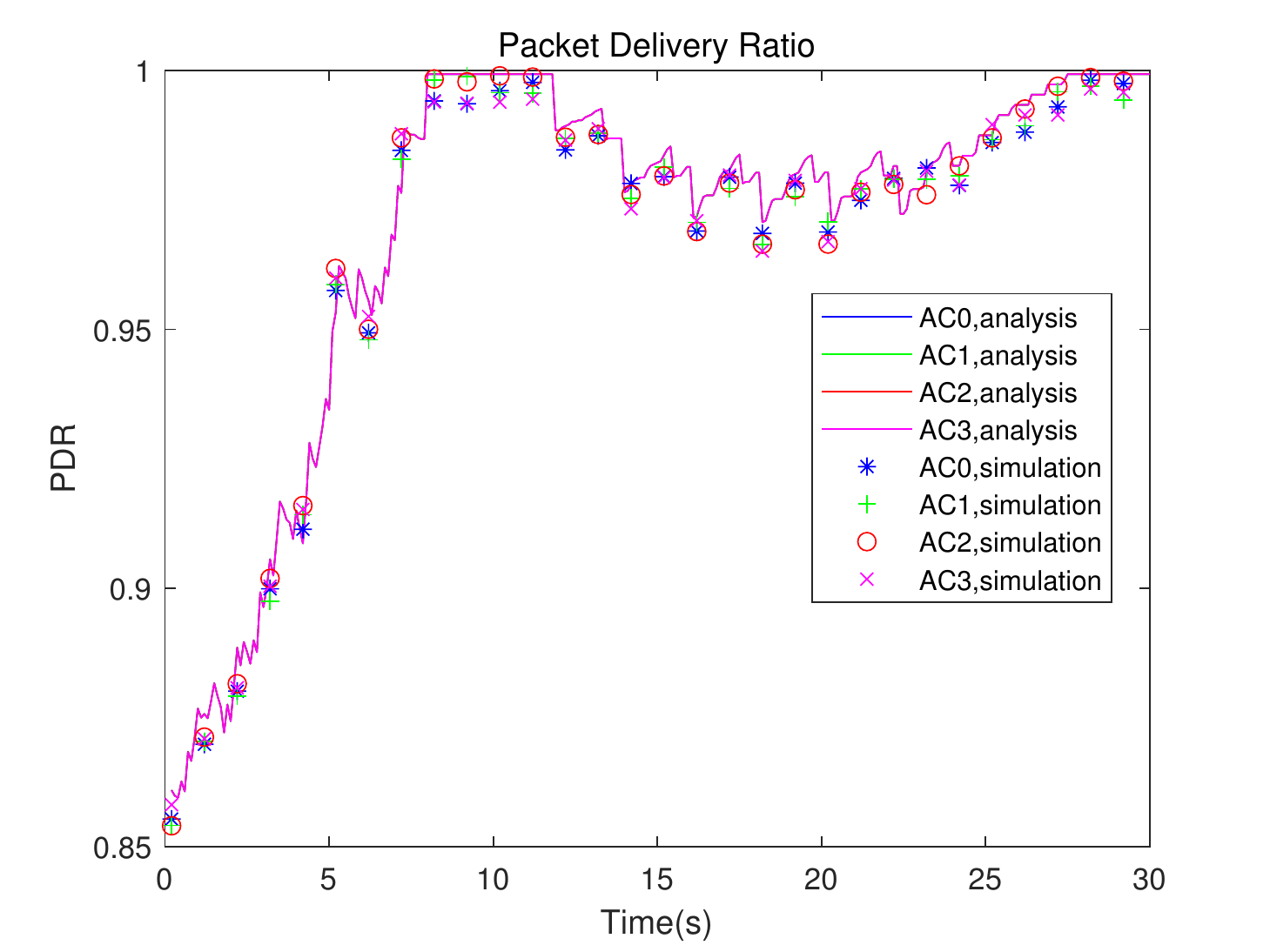}
    \end{minipage}}
  \subfigure[]{
    \begin{minipage}{5.7cm}
    \includegraphics[scale=0.44]{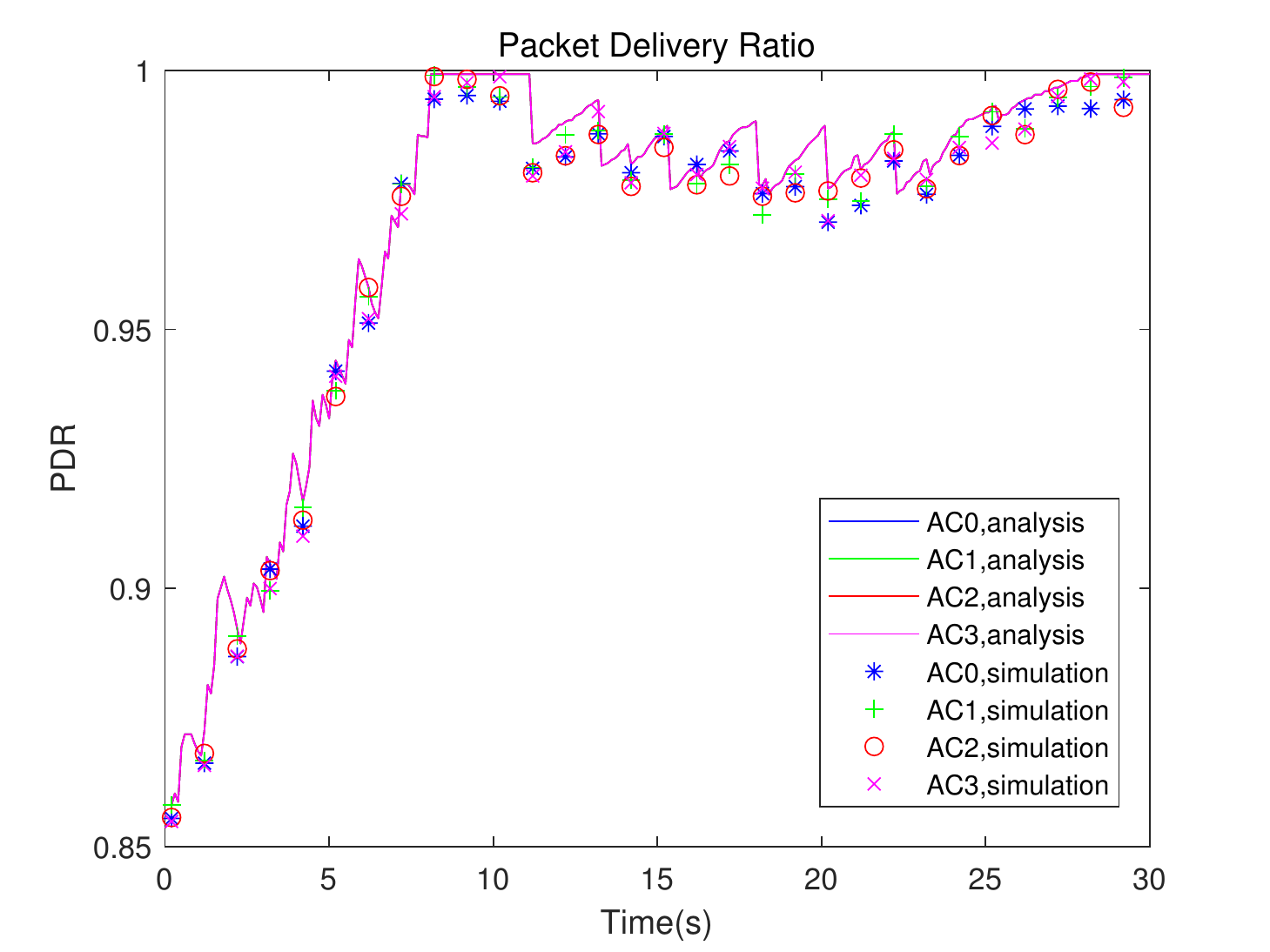}
    \end{minipage}}
  \subfigure[]{
    \begin{minipage}{5.7cm}
    \includegraphics[scale=0.44]{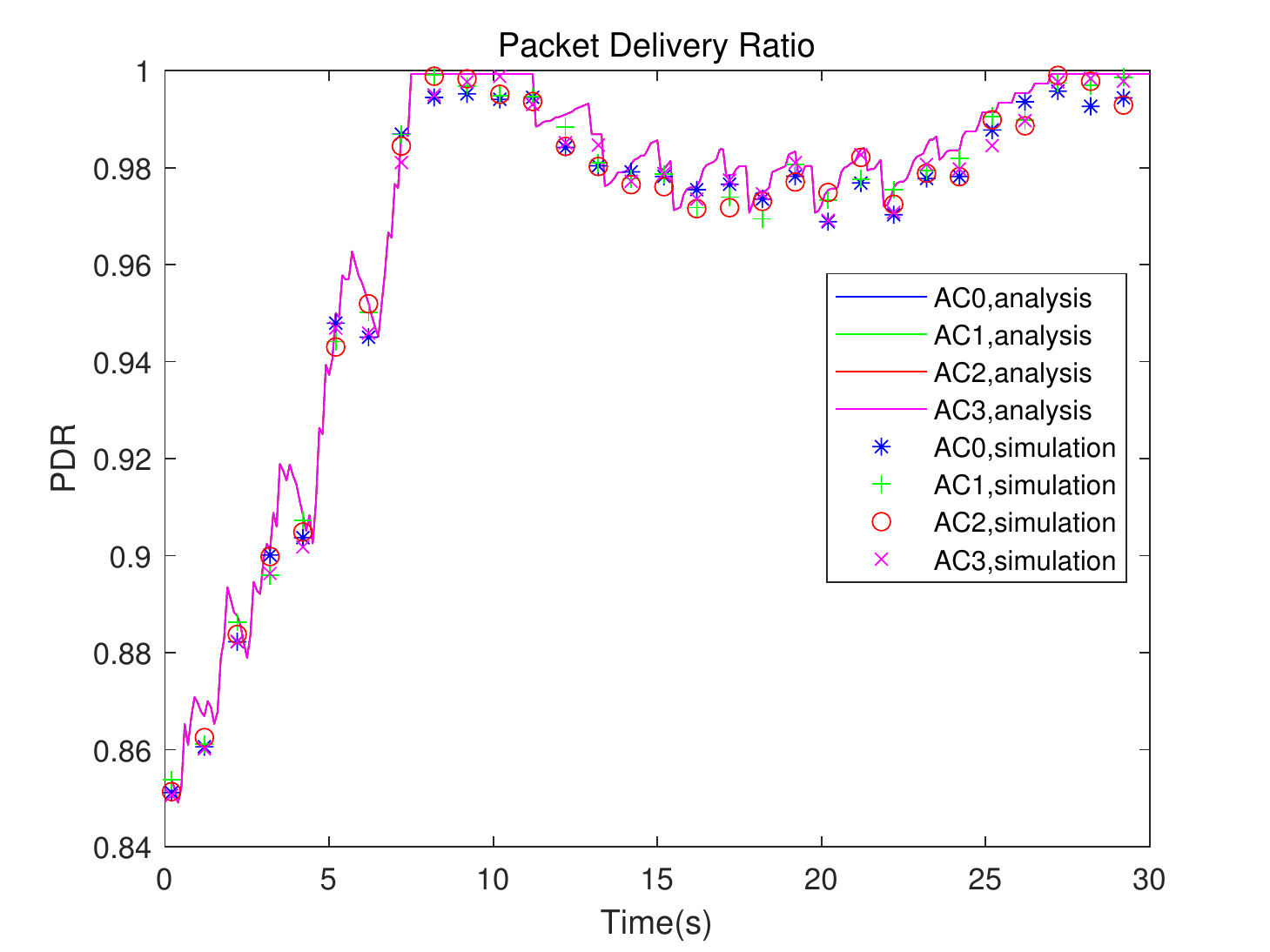}
    \end{minipage}}
    \caption{The time-dependent packet delivery ratio. (a) Left-turn; (b) Straight; (c) Right-turn.}
    \label{fig7}
    \vspace{-0.7cm}
\end{figure*}


Fig. \ref{fig4} shows the time-dependent number of vehicles in the communication range of $V_{1,1}$ (i.e., $N_{c}^{k,i}(t)$ in Eq. \eqref{eq35}), when it initially drives on different lanes. It can be observed that the time-dependent $N_{c}^{1,1}(t)$ always increases first, then decreases, and finally keeps stable, which is independent of its initial lane. This is because platoons first enter the intersection from different directions, resulting in the increasing number of vehicles in the communication range of $V_{1,1}$. Since the communication range is much larger than the intersection, $N_{c}^{1,1}(t)$ will keep stable for a period of time after all the vehicles enter the intersection. Then the platoons that can pass the stop line within the remaining green light time will gradually drive away from the intersection, and thus $N_{c}^{1,1}(t)$ decreases accordingly. Finally, the platoons driving away from the intersection exit the simulation scenario while other platoons stop at the intersection, thus $N_{c}^{1,1}(t)$ keeps stable.

Figs. \ref{fig5}-(a), (b) and (c) show the time-dependent mean service time of four ACs, i.e., $T_{k,i,m}(t)$ in Eq. \eqref{eq41}, when $V_{1,1}$ initially drives on different lanes. It can be seen that the simulation results are very close to the analytical results, which indicates that the analytical model is accurate. Moreover, it is seen that the trend of each AC's service time is consistent with that of $N_{c}^{1,1}(t)$. It is because that the collision probability caused by the exposed vehicles and hidden vehicles increases with $N_{c}^{1,1}(t)$, and thus service time increases, which matches Eq. \eqref{eq34}. In addition, we can find that the AC with lower priority will lead to a longer service time. This is because the contention window size of the AC with lower priority is larger, and thus the AC has a longer backoff time, incurring a longer service time.

Figs. \ref{fig6}-(a), (b) and (c) show the time-dependent packet transmission delay of four ACs, i.e., $PTD_{k,i,m}(t)$ calculated by Eq. \eqref{eq45}, when $V_{1,1}$ initially drives on different lanes, respectively. It can be seen that the simulation results are very close to the analytical results. Moreover, it can be seen that the trend of each AC's time-dependent packet transmission delay is consistent with that of $N_{c}^{1,1}(t)$. This is because that $PTD_{k,i,m}(t)$ is determined by $N_{c}^{k,i}(t)$ according to Eqs. \eqref{eq27}-\eqref{eq42}. In addition, it is shown that the maximum time-dependent packet delays of each AC is smaller than the maximum delay limit, i.e., $100 ms$, which indicates that packets can be received by other vehicles in time.

Figs. \ref{fig7}-(a), (b) and (c) show the time-dependent packet delivery ratio of four ACs, i.e., $PDR_{k,i,m}(t)$ calculated by Eq. \eqref{eq46}, when $V_{1,1}$ initially drives on different lanes, respectively. It is found that the simulation results are almost match the analytical results. Moreover, we can see that the packet delivery ratio first increases and then the packet delivery ratio reaches the maximum value and keeps for a short period. This is because that the number of hidden vehicles of $V_{1,1}$ decreases when platoons enter the intersection, which results in the increasing packet delivery ratio. Later, the target vehicle has no hidden terminal when all vehicles in the platoon enter the communication range of $V_{1,1}$. Afterward, the packet delivery ratio first decreases, then increases, and finally keeps constant. It is because that several platoons will first leave the intersection within the remaining green light time and the hidden vehicles of $V_{1,1}$ appear and gradually increase. When the platoons are far away enough from the other platoons stopping at the intersection and exiting the communication range of $V_{1,1}$, the number of hidden vehicles of $V_{1,1}$ will gradually decrease until there is no hidden vehicle. In addition, the time-dependent packet delivery ratio of each AC is high under different initial scenarios, which means that vehicles can receive most packets successfully.

\section{Conclusions}
\label{sec7}
In this paper, we have constructed time-dependent models of the 802.11p for platooning communications at intersection. We first considered the movement behaviors of platoons at the intersection including turning, accelerating, decelerating and stopping as well as the periodic change of traffic lights to construct the movement model of the vehicles, and then established the hearing network to reflect the time-varying connectivity among vehicles. After that, we adopted PSFFA to model the dynamic behavior of transmission queue, then considered the continuous backoff freezing and adopted the z-domain linear model to describe the access process of the 802.11p with four ACs. Finally, we derived the PTD and PDR based on the time-dependent model of the 802.11p. The accuracy of the analytical model was demonstrated by comparing the simulation results and analytical results. According to the analytical results, we can evaluate the PTD and PDR of 802.11p based on the time-dependent model and draw the following conclusions:

 \begin{itemize}
 \item Since the maximum time-dependent packet transmission delays of each AC is smaller than the maximum delay limit, i.e., $100ms$, the 802.11p can meet the requirement of platooning communications at intersection.

\item The time-dependent packet transmission delay is primarily affected by the number of vehicles in the communication range of the target vehicle, and the time-dependent packet delivery ratio is primarily affected by the hidden vehicles.
\end{itemize}

\section*{Acknowledgment}
The authors are also indebted to Mr. Siyang Xia with Nanjing University, Prof. Pingyi Fan with Tsinghua University and Prof. Jiangzhou Wang with University of Kent, for their help with this work.

\ifCLASSOPTIONcaptionsoff
  \newpage
\fi


\bibliographystyle{IEEEtran}

\begin{IEEEbiography}[{\includegraphics[width=1in,height=1.25in,clip,keepaspectratio]{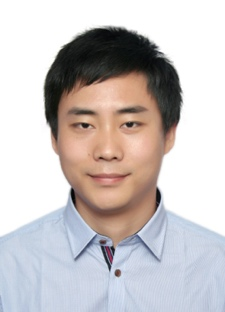}}]{Qiong Wu} (Member, IEEE) received the Ph.D. degree in information and communication engineering from the National Mobile Communications Research Laboratory, Southeast University, Nanjing, China, in 2016.

From 2018 to 2020, he was a Postdoctoral Researcher with the Department of Electronic Engineering, Tsinghua University. He is currently an Associate Professor with the School of Internet of Things Engineering, Jiangnan University, Wuxi, China. His current research interest focuses autonomous driving communication technology.
\end{IEEEbiography}

\begin{IEEEbiography}
[{\includegraphics[width=1in,height=1.25in,clip,keepaspectratio]{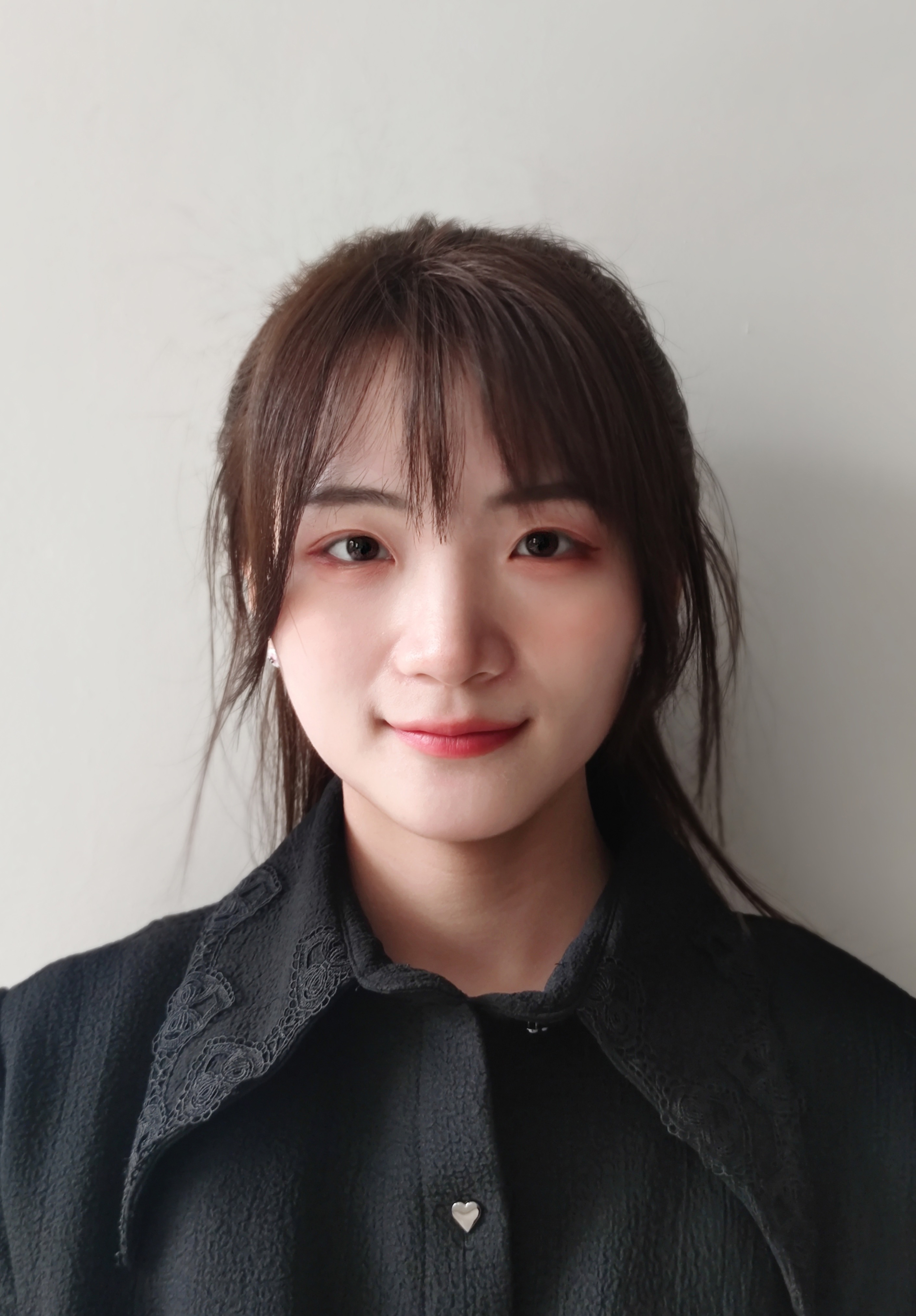}}]
{Yu Zhao} received the B.S. degree from Zhengzhou University of Light Industry, Zhengzhou, China, in 2020. She is currently working toward the M.S. degree at Jiangnan University. Her current research interests include federated learning, deep learning and mobile edge computing.
\end{IEEEbiography}

\begin{IEEEbiography}[{\includegraphics[width=1in,height=1.25in,clip,keepaspectratio]{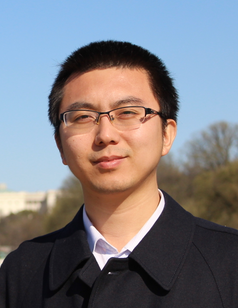}}]{Qiang Fan} received his Ph.D. degree in Electrical and Computer Engineering from New Jersey Institute of Technology (NJIT) in 2019, and his M.S. degree in Electrical Engineering from Yunnan University of Nationalities, China, in 2013. He was a postdoctor researcher in the Department of Electrical and Computer Engineering, Virginia Tech. Currently, he is a staff engineer in Qualcomm, USA. He has served as a reviewer for over 120 journal submissions such as IEEE Transactions on Cloud Computing, IEEE Journal on Selected Areas in Communications, IEEE Transactions on Communications. His current research interests include wireless communications and networking, mobile edge computing, machine learning and drone assisted networking.
\end{IEEEbiography}

\end{document}